\documentclass[10pt,
prd,
twocolumn,
nofootinbib,
noshowkeys,
noshowpacs,
superscriptaddress,
floatfix
]{revtex4-2}
\usepackage{amsmath,amsfonts,amsthm,amssymb}
\usepackage[dvips]{graphics,graphicx}
\usepackage[usenames,dvipsnames]{color}
\definecolor{darkblue}{RGB}{0,0,196}
\definecolor{darkgreen}{RGB}{0,120,0}
\usepackage[colorlinks=true,linktocpage=true,linkcolor=darkblue,citecolor=red,urlcolor=darkblue]{hyperref}
\usepackage{cancel}
\usepackage{bbold}
\usepackage{subfigure}\usepackage{stix}
\usepackage{multirow}
\usepackage{longtable}
\usepackage{color}
\usepackage[normalem]{ulem}
\usepackage{hyperref}
\usepackage{bigints}
\usepackage{xparse}
\usepackage{physics}
\usepackage{verbatim}
\usepackage{minibox}
\usepackage{comment}
\usepackage{appendix}
\usepackage{scalerel}
\usepackage{marginnote}
\usepackage{graphicx}
\usepackage[nice]{nicefrac}
\usepackage{soul}

\usepackage{stackengine}

\newcommand\oU[1]{\ensurestackMath{\stackon[1pt]{#1}{\mkern2mu\bullet}}}
\newcommand\oX[1]{\ensurestackMath{\stackon[1pt]{#1}{\mkern2mu\star}}}
\newcommand\oY[1]{\ensurestackMath{\stackon[1pt]{#1}{\mkern2mu\smwhitestar}}}
\newcommand\oZ[1]{\ensurestackMath{\stackon[1pt]{#1}{\mkern2mu\scaleto{\circ}{3pt}}}}
\def\HP{\hphantom{\alpha}} 
\usepackage{amsmath}
\usepackage{hepunits}

\def\be{\begin{equation}}
	\def\ee{\end{equation}}
	\newcommand{\eel}{\end{eqnarray}}
\def\barr{\begin{array}}
	\def\earr{\end{array}}
\def\beq{\begin{eqnarray}}
	\def\eeq{\end{eqnarray}}
\def\bfig{\begin{figure}}
	\def\efig{\end{figure}}
\newcommand{\bea}{\begin{eqnarray}}
	\newcommand{\eea}{\end{eqnarray}}

\def\lt{\left}
\def\rt{\right}

\newcommand{\nn}{\nonumber}
\newcommand{\f}[2]{\frac{#1}{#2}}

\newcommand{\p}{\partial}

\newcommand{\rf}[1]{Eq.~(\ref{#1})}

\newcommand{\rfmtwo}[2]{Eqs.~(\ref{#1})-(\ref{#2})}
\newcommand{\rfn}[1]{(\ref{#1})}

\def\a{\alpha}
\def\b{\beta}
\def\g{\gamma}
\def\d{\delta}

\def\LR{\left(} 
\def\RR{\right)}
\def\LS{\left[} 
\def\RS{\right]}

\def\HP{\hphantom{\alpha}} 

\newcommand{\sh}[1]{\sinh#1}
\newcommand{\ch}[1]{\cosh#1}

\newcommand{\tU}{\theta_U}
\newcommand{\tX}{\theta_X}
\newcommand{\tY}{\theta_Y}
\newcommand{\tZ}{\theta_Z}

\def\half{\frac{1}{2}}

\newcommand{\lab}[1]{\label{#1}}
\def\nn{\nonumber}

\def\pv{{\boldsymbol p}}
\def\av{{\boldsymbol a}}
\def\bv{{\boldsymbol b}}

\def\be{\begin{equation}}
\def\ee{\end{equation}}
\def\ba{\begin{eqnarray}}
\def\ea{\end{eqnarray}}   

\def\a{\alpha}
\def\b{\beta}
\def\g{\gamma}
\def\d{\delta}

\def\LR{\left(} 
\def\RR{\right)}
\def\LS{\left[} 
\def\RS{\right]}

\def\half{\frac{1}{2}}

\def\pv{{\boldsymbol p}}
\def\av{{\boldsymbol a}}
\def\bv{{\boldsymbol b}}

\def\half{\frac{1}{2}}

\def\n0{n_{(0)}}
\def\e0{\varepsilon_{(0)}}
\def\P0{P_{(0)}}

\def\uv{{\boldsymbol U}}

\def\ov{{\boldsymbol \omega}}

\def\ev{{\boldsymbol e}}
\def\bv{{\boldsymbol b}}

\newcommand{\UD}[1]{\oU{#1}}
\newcommand{\XD}[1]{\oX{#1}}
\newcommand{\YD}[1]{\oY{#1}}
\newcommand{\ZD}[1]{\oZ{#1}}

\def\pv{{\boldsymbol p}}
\begin{document}
\preprint{}
 
    \title{Spin polarization dynamics in the non-boost-invariant background} 
    \author{Wojciech Florkowski}
\email{wojciech.florkowski@uj.edu.pl}
	\affiliation{Institute of Theoretical Physics, Jagiellonian University, PL-30-348 Krak\'ow, Poland}
	\author{Radoslaw Ryblewski}
\email{radoslaw.ryblewski@ifj.edu.pl}
	\author{Rajeev Singh}
\email{rajeev.singh@ifj.edu.pl}
%
%
\author{Gabriel Sophys}
\email{gabriel.sophys@ifj.edu.pl}
%
\affiliation{Institute of Nuclear Physics Polish Academy of Sciences, PL 31-342 Krak\'ow, Poland}%
	\date{\today} 
	\bigskip
\begin{abstract}
Space-time evolution of spin polarization within the framework of hydrodynamics with spin based on de Groot - van Leeuwen - van Weert forms of energy-momentum and spin tensors is studied. Due to the non-boost invariant flow in the system the spin polarization components couple to each other implying some effects on the spin polarization observables.
We study transverse-momentum and rapidity dependence of mean spin polarization vector for $\Lambda$ hyperons. Our results show qualitative agreement for rapidity dependence of the global spin polarization with the experiments and other models. The quadrupole structure of the longitudinal component at midrapidity is not found, however, as compared to the results for Bjorken expansion, some non-trivial signal at forward rapidities is observed.
\end{abstract}
     
\date{\today}


	
\maketitle
%
%
\section{Introduction}
\label{sec:introduction} 
%
In the last two decades, relativistic hydrodynamics has become a well established theory with broad applications in relativistic heavy-ion collisions, condensed matter physics, and astrophysics~\cite{Andersson:2006nr,Florkowski:1321594,rezzolla2013relativistic,Gale:2013da,Jeon:2015dfa,Jaiswal:2016hex}. This, in turn, has paved the way to further extensions of the standard hydrodynamic formalism~\cite{Florkowski:2017olj,Romatschke:2017ejr}, as well as allowed for determining some of the quark-gluon plasma properties~\cite{Busza:2018rrf,Schenke:2021mxx}.  

\smallskip
The recent measurements of spin polarization of particles produced in relativistic heavy-ion collisions~\cite{HADES:2014ttv,STAR:2017ckg,Adam:2018ivw,Niida:2018hfw,STAR:2019erd,ALICE:2019aid,Acharya:2019ryw,Kornas:2019,ALICE:2019onw,STAR:2021beb} gave a new perspective to these studies~\cite{Becattini:2020ngo,Becattini:2021lfq}. 
In particular, the idea of local thermodynamic equilibrium with spin degrees of freedom has been proposed~\cite{Becattini:2007sr,Becattini:2013fla} and has been shown to explain some of the spin polarization phenomena, such as collision energy dependence, in terms of the so called polarization-vorticity coupling~\cite{Becattini:2016gvu,Karpenko:2016jyx,Pang:2016igs,Xie:2017upb,Ambrus:2020oiw}. However, after some initial successes, the spin-thermal models have been proven being unsuccessful in explaining various more differential observables. In particular, the measured transverse-momentum dependence of the spin polarization along the beam direction~\cite{STAR:2019erd,ALICE:2021pzu} has been shown to have an opposite sign with respect to the model predictions~\cite{Becattini:2017gcx,Florkowski:2019voj}. This mismatch between theory and experiment, currently known as the sign problem,  has triggered further theoretical developments raising questions of spin non-equilibrium effects and its genuine dynamics. Among others, the idea of incorporating spin degrees of freedom in the hydrodynamic framework has gained a significant attention as it opened the possibility of probing purely quantum features of the matter in the classical hydrodynamic framework~\cite{Florkowski:2018fap,Speranza:2020ilk,Bhadury:2021oat}. 
\smallskip

The formulation of relativistic hydrodynamics with spin based on quantum kinetic theory was first proposed in Ref.~\cite{Florkowski:2017ruc} and further developed in Refs.~\cite{Florkowski:2017dyn,Florkowski:2018ahw,Becattini:2018duy,Florkowski:2019qdp,Singh:2020rht,Bhadury:2020puc,Bhadury:2020cop,Shi:2020htn,Singh:2021man,She:2021lhe}. Other interesting approaches have used the methods of the effective action~\cite{Montenegro:2018bcf,Montenegro:2020paq,Gallegos:2021bzp}, entropy current analysis~\cite{Hattori:2019lfp,Fukushima:2020ucl,Li:2020eon}, holographic methods~\cite{Gallegos:2020otk,Garbiso:2020puw} and non-local collisions~\cite{Hidaka:2018ekt,Yang:2020hri,Wang:2020pej,Weickgenannt:2020aaf,Weickgenannt:2021cuo,Sheng:2021kfc}.

\smallskip
On general grounds, the dynamics of spin polarization is expected to be controlled by a rank-two anti-symmetric tensor $\omega^{\alpha\beta}$ known as the spin polarization tensor~\cite{Florkowski:2017ruc,Florkowski:2017dyn, Florkowski:2018ahw}. It introduces into hydrodynamics six extra Lagrange multipliers which, together with the standard ones, must be determined based on the conservation laws. The spin polarization tensor is, in general, independent of the so-called thermal vorticity which plays a central role in the spin-thermal models~\cite{Becattini:2013fla,Becattini:2013vja,Becattini:2016gvu,Karpenko:2016jyx,Sun:2017xhx,Li:2017slc,Becattini:2017gcx,Wei:2018zfb,Xia:2018tes,Sun:2018bjl,Gao:2019znl,Ivanov:2019wzg,Kapusta:2019ktm,Gao:2020lxh,Deng:2020ygd,Rindori:2021quq,Liu:2021uhn,Fu:2021pok,She:2021lhe,Gao:2021rom,Yi:2021ryh,Peng:2021ago,Ryu:2021lnx,Wang:2021ngp,Hongo:2021ona,Becattini:2021iol,Becattini:2021suc,Dong:2021fxn}.

The formalism of hydrodynamics with spin presented in Refs~\cite{Florkowski:2017ruc,Florkowski:2017dyn} is based on the forms of the energy-momentum and spin tensors introduced by de Groot, van Leeuwen, and van Weert~\cite{DeGroot:1980dk}, which have been shown to be connected to canonical expressions (obtained by applying Noether theorem) through pseudo-gauge transformations~\cite{Hehl:1976vr,Florkowski:2018ahw}. Following the experimental results which show small values of polarization in Refs.~\cite{Florkowski:2017ruc,Florkowski:2017dyn} it has been always assumed that $\omega^{\alpha\beta}$
is of leading order, hence making the spin polarization contributions appearing only in the spin tensor. The first applications of this framework demonstrating the connection between the theoretical calculations and the experimental measurements have been presented in Ref.~\cite{Florkowski:2019qdp} 

In the current work we extend the analysis of the space-time evolution of the spin degrees of freedom within the framework of  perfect-fluid hydrodynamics with spin previously performed for Bjorken-expanding system~\cite{Florkowski:2019qdp}.
We explicitly break the boost-invariance in the beam direction while keeping the assumption of transverse homogeneity intact. In order to keep our analysis simple and clear, we consider an ideal relativistic gas of classical massive particles of single species with spin-$\frac{1}{2}$~\cite{DeGroot:1980dk,Florkowski:1321594}. 
Relaxing the assumption of boost-invariance leads to non-trivial effects resulting from the longitudinal expansion of the system as well as introduces mixing between different electric-like and magnetic-like sectors of spin coefficients. Using a physics motivated choice of the initial condition for background and spin variables we evolve the system until freeze-out and subsequently analyse the impact of the dynamics on the spin polarization observables. In particular, we study the transverse momentum and azimuthal angle dependence of the mean spin polarization vector as well as its single-differential rapidity dependence. Our results show some interesting patterns in rapidity which are in agreement with other studies and may be potentially important for the spin polarization measurements~\cite{STAR:2018gyt,xu:2021imi,ALICE:2021pzu}. 

The paper is organized as follows: We start with the conventions used in this article in Sec.~\ref{sec:convention} followed by a brief review of perfect-fluid spin hydrodynamics in Sec.~\ref{sec:spinhydro}. After describing our setup in Sec.~\ref{sec:NBIflow}, we derive evolution equations for background and spin in Sec.~\ref{sec:NBI}. Our numerical results for the perfect-fluid background and spin dynamics are presented in Sec.~\ref{sec:numerics} and followed by the analysis on spin polarization observables in Sec.~\ref{sec:spinpolparticle}.
We summarize our findings in Sec.~\ref{sec:summ}.
%
\section{Conventions}
\label{sec:convention}
%
In this paper, the metric tensor is taken with the ``mostly minus'' convention,  $g_{\alpha\beta} =  \hbox{diag}(+1,-1,-1,-1)$, whereas the scalar (or dot) product of two four-vectors $a^{\alpha}$ and $b^{\alpha}$ reads $a \cdot b =a^{\alpha}b_{\alpha}= g_{\alpha \beta} a^\alpha b^\beta = a^0 b^0 - \av \cdot \bv$, where three-vectors are denoted by bold font. For the  Levi-Civita tensor $\epsilon^{\alpha\beta\gamma\delta}$, we use the convention $\epsilon^{0123} = +1$. The Lorentz-invariant measure in the momentum space is given by $dP = d^3p/(E_p(2 \pi )^3)$, where on-mass-shell particle energy and the particle four-momentum are $E_p = \sqrt{m^2 + \pv^2}$ and $p^\mu = (E_p, \pv)$, respectively. We use a shorthand notation for anti-symmetrization by a pair of square brackets. For example, for a rank-two covariant tensor $M$,
one has $M_{[\mu \nu]} = \frac{1}{2}\left(M_{\mu\nu} - M_{\nu\mu} \right)$. The Hodge dual of any tensor ${C}^{\alpha\beta}$ is  denoted by a tilde and  obtained by contracting the rank-two anti-symmetric tensor with the Levi-Civita tensor, namely
\ba
\widetilde{C}^{\alpha\beta} = \f{1}{2} \epsilon^{\alpha\beta\gamma\delta}  {C}_{\gamma\delta}\,.
\lab{eq:dual}
\nonumber
\ea
We also use the following shorthand notation for the directional derivatives
$U^\alpha \partial_\alpha   \equiv \UD{(\phantom{x})}$,
$X^\alpha \partial_\alpha  \equiv \XD{(\phantom{x})}$,
$Y^\alpha \partial_\alpha  \equiv \YD{(\phantom{x})}$,
$Z^\alpha \partial_\alpha  \equiv \ZD{(\phantom{x})}$,
as well as the for the divergence of a four-vector $A$, 
$\partial_\alpha A^\alpha  \equiv \theta_A$.
Throughout the paper we assume natural units {\it i.e.} $c = \hbar = k_B~=1$.
%

\section{Perfect-fluid spin hydrodynamics}
\label{sec:spinhydro}
%
In this section we review the hydrodynamic framework for spin-$\frac{1}{2}$ particles based on the GLW (de Groot - van Leeuwen - van Weert) forms of energy-momentum and spin tensors. In this formalism, the spin effects are considered being small, implying that the spin degrees of freedom are not included in the conservation laws for charge, energy and linear momentum, and appear only in conservation law for angular momentum~\cite{Florkowski:2018ahw}.
%
\subsection{Conservation of charge}
%
The conservation law for baryon number has the form
\ba
\p_\alpha N^\alpha(x)  = 0\,,
\lab{Ncon}
\ea
where the baryon current is
\ba
N^\alpha = {\cal N} U^\alpha\,,
\lab{Nmu}
\ea
with $U^{\a} = \gamma (1,\boldsymbol{v})$ being the fluid four-velocity and \cite{Florkowski:2017ruc}
\ba
{\cal N} = 4 \, \sinh(\xi)\, {\cal N}_{(0)}(T)\,,
\lab{nden}
\ea
stands for particle density.

In the case of an ideal relativistic gas of classical massive particles (herein referred to as the $\Lambda$ EoS), the number density of spinless and neutral classical massive particles, ${\cal N}_{(0)}(T)$,  has the form
\beq
{\cal N}_{(0)}(T) &=& k T^3 z^2 \, K_2\left( z\right), \label{polden}
\eeq
with $k\equiv\f{1}{2\pi^2}$, $z$ being the ratio of the particle mass $m$ over the temperature $T$, $z\equiv m/T$, and  $K_2$ denoting the modified Bessel function of the second kind.

One should note that the factor $4 \, \sinh(\xi) = 2 \left(e^\xi - e^{-\xi} \right)$ in~\rf{nden} represents spin degeneracy and the presence of both particles and antiparticles in the system with $\xi$ being the ratio of the baryon chemical potential ${\mu}$ over the temperature, $\xi\equiv{\mu}/T$.
%
\subsection{Conservation of energy and linear momentum}
\label{subsec:ConTmunu}
%
The conservation law for energy and linear momentum is 
\ba
\p_\a T^{\a\b}(x) = 0,
\lab{Tcon}
\ea
where $T^{\a\b}$ is the energy-momentum tensor having the perfect-fluid form
\ba
T^{\a\b} &=& ({\cal E}+ {\cal P} ) U^\a U^\b - {\cal P} ~g^{\a\b}
\lab{Tmn}
\ea
with the energy density and pressure written as \cite{Florkowski:2017ruc}
\beq
{\cal E} &=& 4 \, \cosh(\xi) \, {\cal E}_{(0)}(T),
\lab{enden}\\
{\cal P} &=& 4 \, \cosh(\xi) \, {\cal P}_{(0)}(T),
\lab{prs}
\eeq
respectively.

Similarly to the number density, the auxiliary energy density, ${\cal E}_{(0)}(T)$, and pressure,  ${\cal P}_{(0)}(T)$, for $\Lambda$ EoS are defined as~\cite{Florkowski:1321594}
\beq
{\cal E}_{(0)}(T) &=& k T^4 \, z ^2  \, 
 \left[z  K_{1} \left( z  \right) + 3 K_{2}\left( z \right) \right],  \label{eneden} \\
{\cal P}_{(0)}(T) &=& k T^4 \, z^2 \,  K_2\left( z\right) = T {\cal N}_{(0)}(T), \label{P0}
\eeq
respectively. Equations~(\ref{Ncon}) and (\ref{Tcon}) together form a closed system of five partial differential equations for five unknown functions: ${\mu}$, $T$, and three independent components of $U^\mu$ (note that $U$ is normalized to $1$). We solve these perfect-fluid equations in order to determine the hydrodynamic background for the spin evolution.
%
\subsection{Conservation of angular momentum}
\label{subsec:ConAM}
%
Since in the GLW formalism the energy-momentum tensor (\ref{Tmn}) is symmetric, the total angular momentum conservation $\partial_{\alpha} J^{\alpha, \beta \gamma}=T^{\beta \gamma}-T^{\gamma \beta}+\partial_{\alpha} S^{\alpha, \beta \gamma}=0$ implies separate conservation of the spin part~\cite{Florkowski:2018ahw}
\beq
\p_\a S^{\a , \beta \gamma }(x)&=& 0\,,
\label{eq:SGLWcon}
\eeq
where, in the leading order of spin polarization tensor $\omega_{\mu\nu}$ (to be discussed later), spin tensor is written as~\cite{Florkowski:2018ahw}
\beq
S^{\alpha , \beta \gamma }
&=&  S^{\alpha , \beta \gamma }_{\rm ph} + S^{\a, \b\g}_{\Delta},
\label{eq:SGLW}
\eeq
with $S^{\alpha , \beta \gamma }_{\rm ph}$\footnote{In Eq.~\rfn{eq:SGLW}, the first term is known as the phenomenological spin tensor~\cite{Florkowski:2017ruc}.}
and $S^{\a, \b\g}_{\Delta}$ defined as
\beq
S^{\alpha , \beta \gamma }_{\rm ph}
&=&  \ch(\xi){\cal N}_{(0)} U^\alpha \omega^{\beta\gamma},\label{Spheno}\\
S^{\a, \b\g}_{\Delta} 
&=&  \ch(\xi)\Big[{\cal A}_{(0)} \, U^\a U^\d U^{[\b} \omega^{\g]}_{\HP\d} \lab{SDeltaGLW} \\
&& \hspace{-0.5cm} + \, {\cal B}_{(0)} \, \Big( 
U^{[\b} \Delta^{\a\d} \omega^{\g]}_{\HP\d}
+ U^\a \Delta^{\d[\b} \omega^{\g]}_{\HP\d}
+ U^\d \Delta^{\a[\b} \omega^{\g]}_{\HP\d}\Big)\Big],
\nn
\eeq
respectively, and thermodynamic coefficients have the forms
\beq 
{\cal B}_{(0)} &=&-\frac{2}{z^2}  \frac{{\cal E}_{(0)}+{\cal P}_{(0)}}{T}\,,
\label{coefB}
\eeq
and
\beq
{\cal A}_{(0)} &=&2 {\cal N}_{(0)}-3{\cal B}_{(0)} .
\label{coefA}
\eeq
The projection operator on the space orthogonal to four-velocity is defined as $\Delta^{\mu}_{\HP \b} = g^{\mu}_{\HP \b} - U^\mu U_\beta$.
%
\subsection{The four-vector basis}
%
For our convenience, in this work we will use the four-vector basis which, apart from the four-velocity $U$, consists 
of additional three four-vectors $X$, $Y$ and $Z$ spanning the space transverse to $U$. The latter are obtained by canonical boost transformation with four-velocity $U$, $\Lambda^\b_\a (U^\mu)$, of the local-rest-frame expressions
\beq
X^\a_{\rm LRF} &=& \Big(0, 1,0, 0\Big),\nn \\ Y^\a_{\rm LRF} &=& \Big(0, 0,1, 0\Big),\label{basislrf}  \\ Z^\a_{\rm LRF} &=& \Big(0, 0,0, 1\Big). \nn
\eeq 
Obviously, the basis vectors satisfy the following conditions
\begin{eqnarray}
 && \,\,U \cdot U = 1,\label{UU}\nonumber \\
X \cdot X \,\,&=& \,\, Y \cdot Y \,\,=\,\, Z \cdot Z \,\,=\,\, -1, \nonumber \\ \label{XXYYZZ}
X \cdot U\,\, &=& \,\,Y \cdot U \,\,=\,\, Z \cdot U \,\,=\,\, 0,  \nonumber \\ \label{XYZU}
X \cdot Y\,\, &=& \,\, Y \cdot Z \,\,=\,\, Z \cdot X \,\,=\,\, 0.  \label{XYYZZX}
\end{eqnarray}
%
\subsection{Spin polarization tensor}
\label{subsec:spinpol}
%
The spin polarization tensor $\omega_{\mu\nu}$ is an anti-symmetric rank-two tensor which can be decomposed in terms of four-vectors $\kappa^\mu$ and $\omega^\mu$ as \cite{Florkowski:2017ruc}
\beq
\omega_{\mu\nu} &=& \kappa_\mu U_\nu - \kappa_\nu U_\mu + \epsilon_{\mu\nu\a\b} U^\a \omega^{\b}. \lab{spinpol1}
\eeq
Note that any part of the four-vectors $\kappa^{\mu}$ and $\omega^{\mu}$ parallel to $U^{\mu}$ does not contribute to the right-hand side of~Eq.~(\ref{spinpol1}).
Therefore, we assume that $\kappa^{\mu}$ and $\omega^{\mu}$ fulfill the following orthogonality conditions~\footnote{Six independent components of $\kappa^{\mu}$ and $\omega^{\mu}$ define six independent components of the anti-symmetric rank-two tensor $\omega_{\mu\nu}$.}
\beq
\kappa\cdot U = 0, \quad \omega \cdot U = 0  \lab{ko_ortho},
\eeq
which allows us to express $\kappa_\mu$ and $\omega_\mu$ in terms of $\omega_{\mu\nu}$ as
\beq
\kappa_\mu= \omega_{\mu\a} U^\a, \quad \omega_\mu = \half \epsilon_{\mu\a\b\gamma} \omega^{\a\b} U^\gamma. \lab{eq:kappaomega}
\eeq
Using the basis vectors and orthogonality conditions \rfn{ko_ortho}, $\kappa^{\mu}$ and $\omega^{\mu}$ can be decomposed in terms of scalar spin coefficients $C_{\boldsymbol{\kappa}} =(C_{\kappa X}, C_{\kappa Y}, C_{\kappa Z})$ and $C_{\boldsymbol{\omega}} =(C_{\omega X}, C_{\omega Y}, C_{\omega Z})$ as follows
\beq
\kappa^\a &=&  C_{\kappa X} X^\a + C_{\kappa Y} Y^\a + C_{\kappa Z} Z^\a, \lab{eq:k_decom}\\
\omega^\a &=&  C_{\omega X} X^\a + C_{\omega Y} Y^\a + C_{\omega Z} Z^\a. \lab{eq:o_decom}
\eeq
Using Eqs.~\rfn{eq:k_decom} and \rfn{eq:o_decom} in \rf{spinpol1} we can obtain a general expression for the spin polarization tensor $\omega_{\alpha\beta}$ in terms of scalar spin coefficients  
\beq
\omega_{\alpha\beta} &=& 2\left(C_{\kappa X} X_{[\alpha} U_{\beta]} + C_{\kappa Y} Y_{[\alpha} U_{\beta]} + C_{\kappa Z} Z_{[\alpha} U_{\beta]}\right) \,\nonumber \\
&& + \epsilon_{\alpha\beta\gamma\delta} U^\gamma \left(C_{\omega X} X^\delta + C_{\omega Y} Y^\delta + C_{\omega Z} Z^\delta\right). 
\label{eq:omegamunu}
\eeq
We may also introduce another parametrization of the spin polarization tensor, which uses electric- and magnetic-like three-vectors in the laboratory frame, $\ev = (e^1,e^2,e^3)$ and $\bv = (b^1,b^2,b^3)$, in the form~\cite{Florkowski:2017dyn}
\begin{equation}
\omega_{\a\b}= 
\begin{bmatrix}
0       &  e^1 & e^2 & e^3 \\
-e^1  &  0    & -b^3 & b^2 \\
-e^2  &  b^3 & 0 & -b^1 \\
-e^3  & -b^2 & b^1 & 0
\end{bmatrix},
\label{omeb}
\end{equation}
where we used the sign conventions of Ref.~\cite{Jackson:1998}. Using \rf{omeb} in \rf{eq:kappaomega} one finds~\cite{Florkowski:2017dyn}
\beq
\kappa^\a &=& (\kappa^0, \boldsymbol{\kappa}) =  \left( \ev \cdot \uv, U^0 \ev + \uv \times \bv \right), \nn \\
\omega^\a &=& (\omega^0, \ov) =  \left( \bv \cdot \uv, U^0 \bv - \uv \times \ev \right) .
\label{komFROMeb}
\eeq
%
\section{Transversely homogeneous system with non-boost-invariant flow}
\label{sec:NBIflow}
%
In the current work we assume that the system is transversely homogeneous and undergoing a non-trivial dynamics along the beam ($z$) direction. Due to translational invariance in the transverse plane, we assume that the fluid four-velocity $U^\alpha$ has vanishing $x$ and $y$ components, namely
\beq
U^\a &=& \Big(\ch(\Phi), 0,0, \sh(\Phi)\Big)~,
\label{flow}
\eeq
where $\Phi=\vartheta+\eta$ is the fluid rapidity with $\vartheta(\tau,\eta)$ being the scalar function of longitudinal proper time $\tau=\sqrt{t^{2}-z^{2}}$ and space-time rapidity $\eta~=~\ln \left[(t+z)/(t-z)\right]/2$ describing the deviations of the flow from the boost-invariant form. Hence, in the limit $\vartheta(\tau,\eta)=0$, we recover the Bjorken  flow~\cite{Bjorken:1982qr}.

In our case basis four-vectors~\rfn{basislrf} take the forms
\beq
X^\a &=& \Big(0, 1,0, 0\Big), \quad Y^\a = \Big(0, 0,1, 0\Big),\nn\\
Z^\a &=& \Big(\sh(\Phi), 0,0, \ch(\Phi)\Big).
\lab{BIbasis}
\eeq 
The directional derivatives read
\beq
U^\alpha \partial_\alpha &=& \cosh({\vartheta}) ~\partial_\tau
 + \frac{\sinh({\vartheta})}{\tau} \partial_\eta   \nonumber \, ,\\
 X^\alpha \partial_\alpha &=&  \partial_x   \nonumber \, ,\\
Y^\alpha \partial_\alpha &=&  \partial_y   \nonumber \, ,\\
Z^\alpha \partial_\alpha &=&  \sinh({\vartheta}) ~\partial_\tau + \frac{\cosh({\vartheta})}{\tau} \partial_\eta
\label{deriv}
\eeq
and the divergences of $U^\alpha$ and $Z^\alpha$ are written as
\beq
\partial_\alpha U^\alpha &=& \frac{\cosh({\vartheta})}{\tau} + \ZD{\vartheta} ~,\nn \\
\partial_\alpha Z^\alpha &=& \frac{\sinh({\vartheta})}{\tau} + \UD{\vartheta}   ~,
\label{div}
\eeq
respectively, where we have used the shorthand notation for operators \rfn{deriv} explained in Sec.~\ref{sec:convention}. 
Note that in our setup, all scalar functions depend on $\tau$ and $\eta$, hence acting on them with $\partial_x$ and $\partial_y$ gives zero.

Within our setup the two parameterizations, (\rf{eq:omegamunu} and \rf{omeb}), are related through the expressions
\beq 
C_{\kappa X} &=& e^1 \ch(\Phi) - b^2 \sh(\Phi),
\label{CKX}\\
C_{\kappa Y} &=& e^2 \ch(\Phi) + b^1 \sh(\Phi),
\label{CKY}\\
C_{\kappa Z} &=& e^3 ,
\label{CKZ}\\
C_{\omega X} &=& b^1 \ch(\Phi) + e^2 \sh(\Phi),
\label{CWX}\\
C_{\omega Y} &=& b^2 \ch(\Phi) - e^1 \sh(\Phi),
\label{CWY}\\
C_{\omega Z} &=& b^3.
\label{CWZ}
\eeq
One can also invert the above relations and write the laboratory-frame spin components in terms of $C_{\boldsymbol{\kappa}}$ and $C_{\boldsymbol{\omega}}$ as
\beq
e^1 &=& C_{\kappa X} \ch(\Phi) + C_{\omega Y} \sh(\Phi),
\label{e1}\\
e^2 &=& C_{\kappa Y} \ch(\Phi) - C_{\omega X} \sh(\Phi),
\label{e2}\\
e^3 &=& C_{\kappa Z},
\label{e3}\\
b^1 &=& C_{\omega X} \ch(\Phi) - C_{\kappa Y} \sh(\Phi),
\label{b1}\\
b^2 &=& C_{\kappa X} \sh(\Phi) + C_{\omega Y} \ch(\Phi) ,
\label{b2}\\
b^3 &=& C_{\omega Z}.
\label{b3}
\eeq
%
\section{Non-boost-invariant forms of the conservation laws}
\label{sec:NBI}
%
Using \rf{Nmu} in \rf{Ncon}, the baryon charge conservation law is expressed as
\beq
\UD{{\cal N}}+{\cal N}~\theta_U=0.
\label{chargecons}
\eeq
Projecting conservation law for energy and linear momentum (\ref{Tcon}) with flow four-vector $U_{\b}$ and transverse projector $\Delta^{\mu}_{\HP \b}$ and using \rf{Tmn}, we get
\beq
\UD{{\cal E}}+({\cal E}+{\cal P})~\theta_U&=&0,\lab{eq:encons}\\
({\cal E}+{\cal P})\UD{U^{\mu}}-\nabla^{\mu} {\cal P}&=&0,\lab{eq:momcons}
\eeq
respectively, where we introduced the notation $\nabla^{\alpha}\equiv\partial^{\alpha}-U^{\alpha} U^\beta \partial_\beta$.
One can see that Eqs.~\rfn{eq:encons} and \rfn{eq:momcons} are relativistic generalizations of continuity and Euler equations, respectively.

To derive evolution equations for the spin coefficients $C_{\boldsymbol{\kappa}}$ and $C_{\boldsymbol{\omega}}$,
it is convenient to have another decomposition of the spin tensor \rfn{eq:SGLW} as~\cite{Florkowski:2019qdp}
\beq
S^{\a, \b\g} 
&=& U^\a \left({\cal A}_1 \omega^{\b\g} 
+ {\cal A}_2 U^{[\b} \kappa^{\g]}\right)
+  {\cal A}_3 \left(U^{[\b} \omega^{\g]\a} + g^{\a[\b} \kappa^{\g]}\right)\nn\\
\lab{SGLW2}
\eeq
where
\beq
{\cal A}_1 &=& \cosh({\xi}) \LR {\cal N}_{(0)} -  {\cal B}_{(0)} \RR \label{A1} ,\\ 
{\cal A}_2 &=& \cosh({\xi}) \LR {\cal A}_{(0)} - 3{\cal B}_{(0)} \RR  \label{A2} , \\ 
{\cal A}_3 &=&  \cosh({\xi})\, {\cal B}_{(0)}\, .
\label{A3}
\eeq
We first put Eqs.~\rfn{eq:k_decom} and \rfn{eq:omegamunu} in \rf{SGLW2}, subsequently we use \rf{SGLW2} in \rf{eq:SGLWcon} and, finally, we contract the final tensorial equation with $U_\b X_\g$, $U_\b Y_\g$, $U_\b Z_\g$,  $Y_\b Z_\g$, $X_\b Z_\g$ and $X_\b Y_\g$, which yields (see App. \ref{app:deriv} for detailed derivation)
\beq
&&\UD{\a_{x1}} + \ZD{\b_{y1}} = - \a_{x1}\tU - \frac{\a_{x2} U\ZD{Z}}{2} - \b_{y1} \tZ + \b_{y2} U\UD{Z}\,,
\label{spineq1}\\
&&\UD{\a_{y1}} - \ZD{\b_{x1}} = - \a_{y1} \tU - \frac{\a_{y2} U\ZD{Z}}{2} + \b_{x1} \tZ - \b_{x2} U\UD{Z}\,,
\label{spineq2}\\
&&\UD{\a_{z1}} = -\a_{z1} \tU\,,
\label{spineq3}\\
&&\frac{\ZD{\a_{y2}}}{2} + \UD{\b_{x2}} = - \frac{\a_{y2} \tZ}{2} + \a_{y1} Z\UD{U} -\b_{x2} \tU - \b_{x1} Z\ZD{U}\,,
\label{spineq4}\\
&&\frac{\ZD{\a_{x2}}}{2}-\UD{\b_{y2}} = - \frac{\a_{x2}\tZ}{2} + \a_{x1}Z\UD{U} + \b_{y2} \tU + \b_{y1} Z\ZD{U}\,,
\label{spineq5}\\
&&\UD{\b_{z2}} = -\b_{z2} \tU\,,
\label{spineq6}
\eeq
respectively. Above we have introduced the quantities\footnote{One may easily obtain the evolution equations for $S^{\a, \b\g}_{\rm ph}$ from \rfmtwo{spineq1}{spineq6} if one sets ${\cal A}_{(0)} = {\cal B}_{(0)} = 0$.}
\beq
\a_{i1} &=&-\Big({\cal A}_1-\frac{{\cal A}_2}{2}-{\cal A}_3\Big) C_{\kappa i}\,,
\label{alpha-i}\\
\a_{i2} &=&-{\cal A}_3 C_{\kappa i}\,,
\label{alpha-i2}\\
\b_{i1} &=&\frac{{\cal A}_3}{2} C_{\omega i}\,,
\label{beta-i1}\\
\b_{i2} &=& {\cal A}_1 C_{\omega i}\,.
\label{beta-i2}
\eeq

Interestingly, according to Eqs.~\rfn{spineq3} and \rfn{spineq6}, evolution of $C_{\kappa Z}$ and $C_{\omega Z}$ decouples from that of other components. This was also the case in our previous study of a system undergoing Bjorken expansion, see Ref.~\cite{Florkowski:2019qdp}. However, unlike in our previous work where each component evolved independently~\cite{Florkowski:2019qdp}, in the current study we find, see Eqs.~\rfn{spineq1} and \rfn{spineq5}, that $C_{\kappa X}$ and $C_{\omega Y}$ mix during the fluid's expansion and affect the evolution of each other. Similarly, from Eqs.~\rfn{spineq2} and \rfn{spineq4}, we observe that $C_{\kappa Y}$ and $C_{\omega X}$ are also coupled to each other and obey the same form of evolution equations as $C_{\kappa X}$ and $C_{\omega Y}$. The latter feature is solely due to assumed rotational invariance in the transverse plane. Since the breaking of the Bjorken symmetry results in the coupling between magnetic-like and electric-like sectors of spin coefficients, the behavior of spin polarization of $\Lambda$ hyperons is more complex as compared to our previous study. 
%
\subsection{Massive limit}
\label{subsec:massive}
%
Due to relatively large masses of $\Lambda$ particles as compared to the temperatures of the system, at top RHIC energies reaching at most $0.5$ GeV in central heavy-ion collisions, the massive limit where $z \gg 1$ is of particular interest.  In this limit the coefficient ${\cal B}_{(0)}\sim T^3  e^{-x} \sqrt{x}$ can be neglected compared to ${\cal N}_{(0)}\sim  T^3 e^{-x} x^{3/2}$ and hence Eq.~\rfn{eq:SGLW} takes the form
\beq
S^{\alpha , \beta \gamma }_{\rm M}
&=& \ch(\xi){\cal N}_{(0)} U^\alpha \epsilon^{\beta\gamma\mu\nu} U_\mu \omega_\nu\,.
\label{Smassive}
\eeq
Moreover, in the massive limit, one finds that only $\b_{i2}$ coefficient \rf{beta-i2} is non-vanishing, hence, Eqs. \rfn{spineq1}-\rfn{spineq3} are satisfied trivially, and the three components of $\omega^\mu$ follow the same differential equation
\beq
\UD{\b_{i2}} = -\b_{i2} \tU\,.
\label{spineqmassive}
\eeq
Clearly, in the limit of heavy constituents the components of $\kappa$ become undetermined, and the spin tensor is entirely defined by three spatial components of $\omega^\mu$. One may also note that Eqs.~\rfn{Smassive} and \rfn{Spheno} are identical if one imposes the Frenkel condition $\kappa^\alpha = 0$ on Eq.~\rfn{Spheno}.

In the case of baryon-free matter, using standard thermodynamic relations and the temperature solutions for Eq.~\rfn{eq:encons}~\cite{Florkowski:1321594} we can write Eq.~\rfn{spineqmassive}  as
\beq
\UD{C}_{\omega X} = - C_{\omega X} \tU \left(1- {c_s^2} \frac{{\cal E}_{(0)}}{{\cal P}_{(0)}}\right)\,,
\label{spineqmassive2}
\eeq
where $c_s^2 =d{\cal P}/d{\cal E}$ is the speed of sound squared. 

Equation \rfn{spineqmassive2} is of high importance as it explains the behavior of the spin coefficients at the edges of the system. The term in the bracket of \rf{spineqmassive2} determines the behavior of the evolution of the spin coefficient at large space-time rapidity where the system enters the region of large $z$. We find that, for the $\Lambda$ EoS it is negative and, in modulus, increasing with $\eta$, hence, the spin coefficients are increasing along the fluid lines which can also be seen in the numerical analysis (see next sections). We note that for general (3+1)D expanding case \rfmtwo{spineq4}{spineq6} pick up additional terms which make the spin components coupled to each other.
%
\section{Numerical results}
\label{sec:numerics}
%
We start the current section by specifying the initial conditions used in our numerical analysis.
Then, using the $\Lambda$ EoS~\cite{DeGroot:1980dk,Florkowski:1321594}, we present the numerical solutions for the evolution of perfect-fluid background and spin coefficients. Finally, we use our solutions to extract the spin polarization observables.
%
\subsection{Initial conditions}
\label{sec:IC}
\begin{figure*}[t]
\centering
\includegraphics[width=8.6cm]{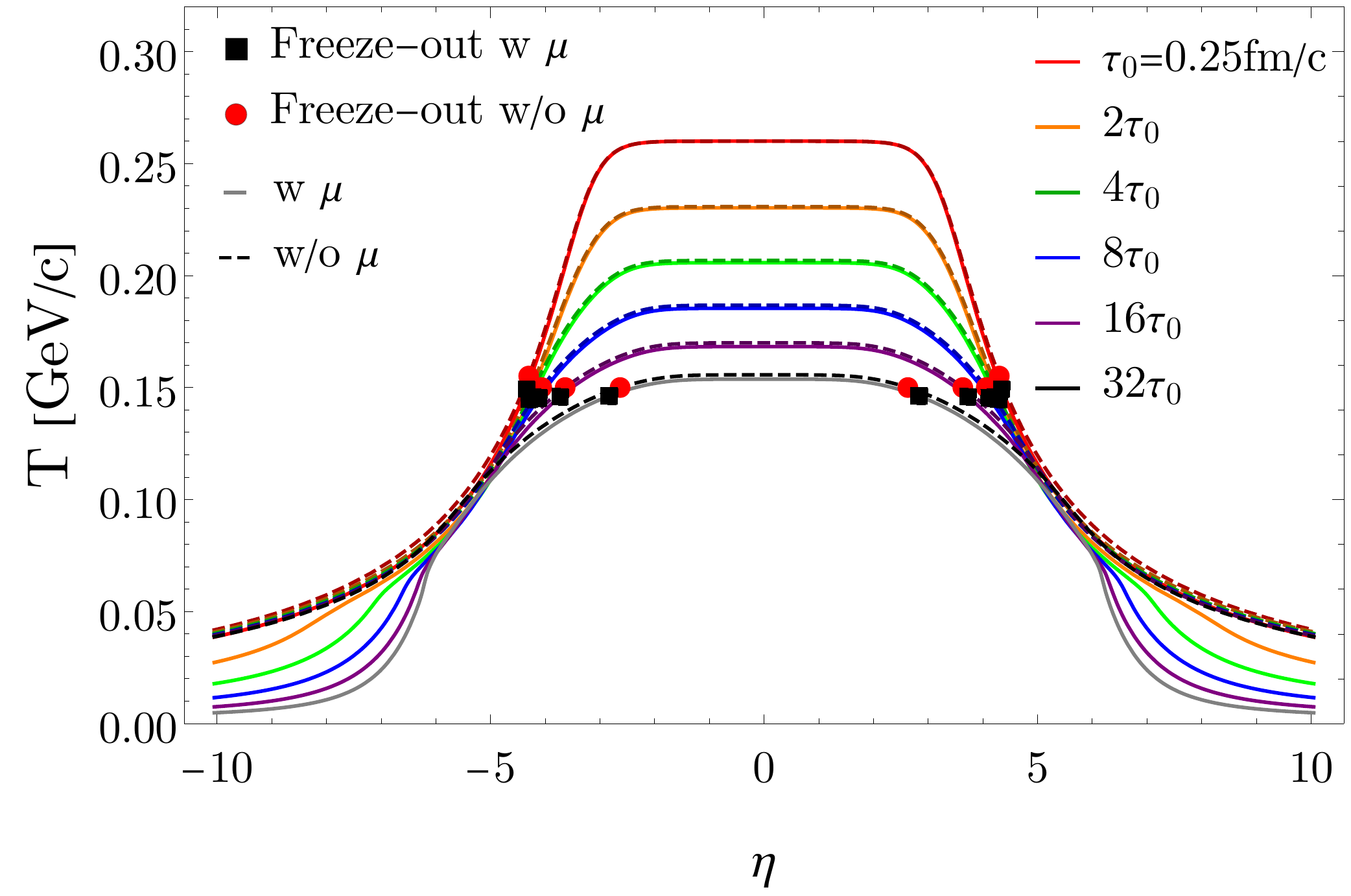}
\includegraphics[width=8.6cm]{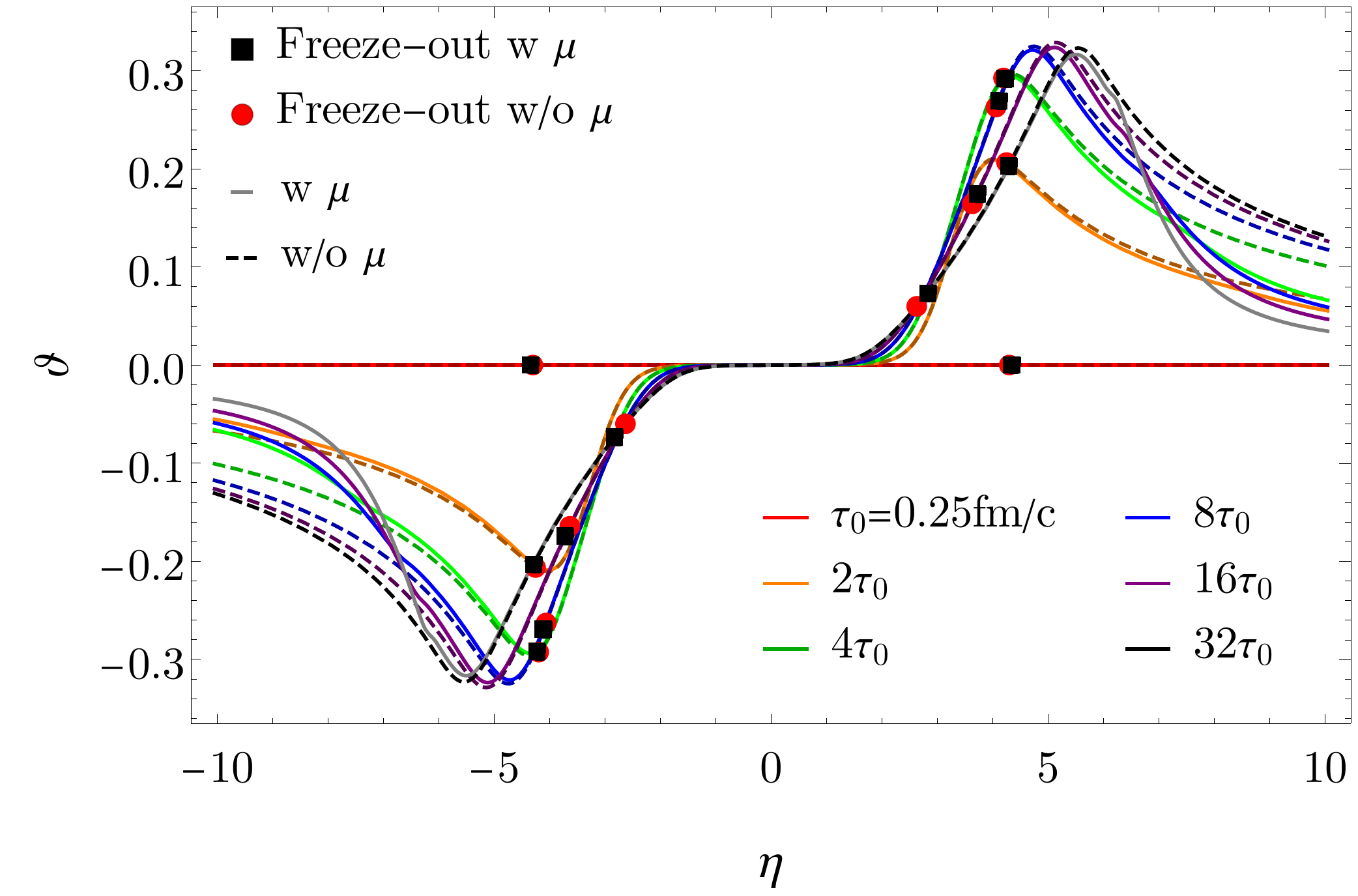}
\caption{Temperature (left) and the deviations of fluid rapidity from Bjorken one (right) in $\eta$ for different longitudinal proper-times: $\tau \in \{1,2,4,8,16,32\}\tau_0 = \{0.25,0.5,1,2,4,8\} {\rm fm}/c$}.
\label{TMu}
\end{figure*}
\begin{figure*}[t]
\centering
\includegraphics[width=8.6cm]{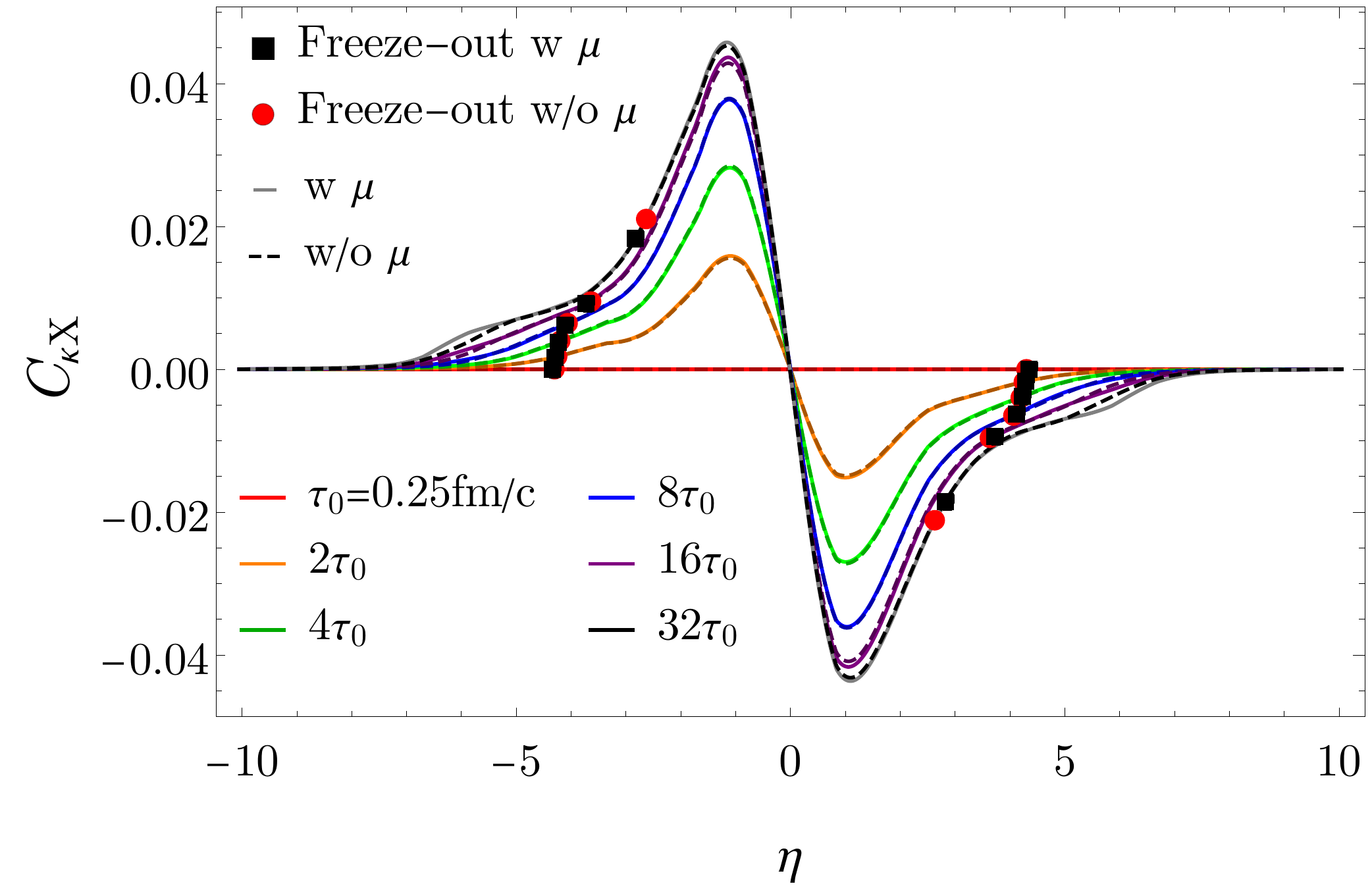}
\includegraphics[width=8.6cm]{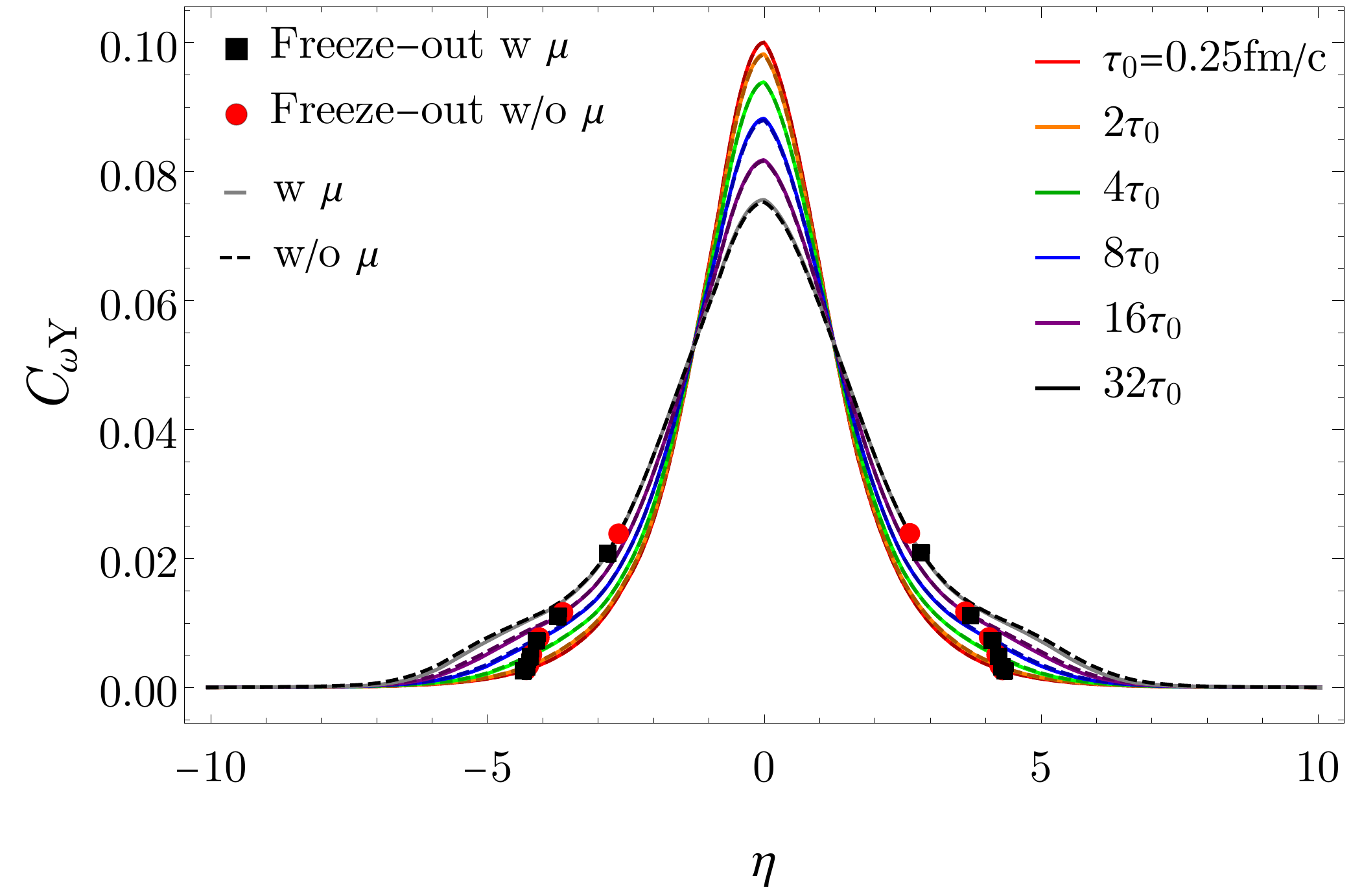}
\caption{Evolution of the spin coefficients $C_{\kappa X}$ (left) and $C_{\omega Y}$ (right) as a function of $\eta$.} 
\label{CkxCwy}
\end{figure*}
We assume initial profile for the energy-density in the form
\beq
{\cal E}_{0} (\eta) &=& \frac{{\cal E}_{0}^c(T_0)}{2} \Bigg[\Theta (\eta ) \Big(\tanh (a-\eta \, b)+1\Big)\nn\\
&&~+ \Theta (-\eta ) \Big(\tanh (a+\eta \, b)+1\Big)\Bigg]
\eeq
where $a=6.2$, $b=1.9$, and ${\cal E}_{0}^c(T_0)$ is the initial central energy density evaluated at the given initial central temperature $T_0 =T(\tau_0,\eta=0) =  260$
MeV, and $\Theta$ is the Heaviside step function~\cite{mabramowitz64:handbook}.

In the current work, we primarily perform the analysis for zero baryon chemical potential aiming at addressing the dynamics of spin at higher collision energies. However, in order to see possible effects of finite baryon chemical potential we also compare our results with the predictions obtained with homogeneous $\mu$ profile choosing $\mu_0=\mu(\tau_0) = 0.12$ GeV. The longitudinal fluid rapidity is chosen to be vanishing initially, $\vartheta_0 (\eta) = 0$.

Initialization of spin coefficients $C_{\boldsymbol{\kappa}}$ and $C_{\boldsymbol{\omega}}$ (or, equivalently, $\ev$ and $\bv$), is performed in such a way as to resemble the physical situation taking place in non-central relativistic heavy-ion collisions. In these processes, the total angular momentum $\boldsymbol{J}$ of the system in the center-of-mass frame has initially only orbital angular momentum part $\boldsymbol{L}$ which is perpendicular to the reaction plane and negative~\cite{STAR:2017ckg,Adam:2018ivw,STAR:2019erd,Acharya:2019ryw,Kornas:2019}. After the collision, due to particle interactions, some fraction of the initial angular momentum is converted to the spin of the system's constituents $\boldsymbol{S}$, namely
\beq
\boldsymbol{J}_{\rm initial} = \boldsymbol{L}_{\rm initial} = \boldsymbol{L}_{\rm final} + \boldsymbol{S}_{\rm final},
\label{eq:L}
\eeq
meaning that, on average, the direction of the initial spin angular momentum is along the original angular momentum~\cite{Florkowski:2019qdp}.

The components of the angular momentum vector of the system are related to the spatial components of the respective angular momentum two-tensor through the relation~\cite{Florkowski:2021pkp}
\begin{equation}
L^k = -\frac{1}{2} \epsilon^{kij} L^{ij},
 \end{equation}
hence the $y$
component of the angular momentum vector is related to non-vanishing $xz$ component of the total angular momentum tensor.

The above physics picture requires only $xz$
component of the spin angular momentum to be non-zero~\cite{Florkowski:2019qdp}, $S^{13}_{\rm FO} \neq 0$ (see App. \ref{app:spincontri} for more details), which is related only to $C_{\kappa X}$ and $C_{\omega Y}$ components of polarization. For the latter, we choose
\beq
C^0_{\kappa X}(\eta)&=&C_{\kappa X}(\tau_0,\eta)= 0\, ,\nn\\ \qquad C^0_{\omega Y}(\eta)&=&C_{\omega Y}(\tau_0,\eta) = d \sech(\eta)\,,
\label{spinini}
\eeq
where we have used $d=0.1$.  

According to Eqs.~\rfn{e1} and \rfn{b2}, components $e^1$ and $b^2$ are related to $C_{\kappa X}$, $C_{\omega Y}$, and fluid rapidity $\vartheta$. Initialization of $C_{\kappa X}$, $C_{\omega Y}$ also initializes $e^1$ and $b^2$
as
\beq
e^1_{0}(\eta)&=& e^1(\tau_0,\eta)= d \tanh(\eta)\, ,\nn\\ \qquad b^2_{0}(\eta)&=& b^2(\tau_0,\eta) = d\,.
\label{spinini2}
\eeq
%
\subsection{Perfect-fluid background evolution}
%
For the perfect-fluid background evolution we solve Eqs.~\rfn{eq:encons}-\rfn{eq:momcons} using $\Lambda$ EoS defined by Eqs.~\rfn{enden}-\rfn{prs}. Figure \rfn{TMu} shows the results for $T$ and $\vartheta$ as a function of $\eta$ for different longitudinal proper times where  $\tau_0 = 0.25$ fm/c with $\mu_0 = 0$ (dashed lines) and $\mu_0 = 0.12$ GeV (solid lines).
 
As expected, we see the symmetric behavior of temperature with respect to the space-time rapidity, $\eta$, in contrast to the behavior of the fluid rapidity. We reproduce well-established result, \textit{i.e.}, temperature decreases with increasing proper time at the center ($\eta = 0$), while, at the same time, the gradients of temperature lead to the build up of the fluid velocity.
One can notice from Fig.~\rfn{TMu} that the temperature and fluid rapidity start to decrease fast at $\eta \approx \pm 5$ eventually decaying to zero at large rapidity. 

In the case of constant initial value of baryon chemical potential the evolution of $T$ and $\vartheta$ undergoes only a mild modification, see Fig.~\ref{TMu}. At the center the behavior of $T$ and $\vartheta$ with respect to the ones with zero baryon chemical potential is similar, with some considerable effect seen only at the edges.
%
\subsection{Spin coefficients evolution}
\label{subsec:spincoeff}
%
An important feature of our spin hydrodynamic framework is that, due to the small polarization limit~\cite{Florkowski:2018ahw,Florkowski:2018fap,Florkowski:2019qdp}  that we work in, the spin evolution does not affect the background evolution meaning that the former may be studied on top of the latter. 

Since the physics scenario considered herein requires only the initialization of $C_{\kappa X}$ and $C_{\omega Y}$ which are coupled to each other through \rf{spineq1} and \rf{spineq5}, we will analyse numerically the evolution of only these two spin coefficients. As can be observed from Fig.~\ref{CkxCwy} the symmetry in $\eta$ of these coefficients remains the same throughout the evolution, which is a feature resulting from the evolution equations (\ref{spineq1}) and (\ref{spineq5}), initial conditions (\ref{spinini}), and  the symmetry of the background.
Due to the coupling between the spin coefficients, $C_{\kappa X}$ is generated during the evolution even if initially chosen equal zero. One may also notice the effect of the temperature evolution on the dynamics of $C_{\omega Y}$ through thermodynamic coefficients, Eqs.~\rfn{coefB} and \rfn{coefA}.
Similarly to the temperature, $C_{\omega Y}$ also decreases with proper time at $\eta = 0$. This behavior of $C_{\omega Y}$ at the center reproduces the case for the Bjorken expanding system.
However, in the region of large rapidity ($\eta \approx \pm 5$) the behavior of the spin is reversed due to the fact that the system reaches the massive limit (see Sec.~\ref{subsec:massive} for extended discussion).

In the case of non-vanishing baryon chemical potential the dynamics of spin coefficients is only slightly affected, see Fig.~\ref{CkxCwy}.

\smallskip
%
\section{Spin polarization of emitted particles}
\label{sec:spinpolparticle}
%
To calculate the spin polarization of particles at freeze-out, we have to define freeze-out hypersurface $ \Sigma_\mu$ by specifying longitudinal proper time $\tau$ for each point of spacetime rapidity $\eta$. Subsequently, we evaluate the average Pauli-Luba\'nski (PL) vector for particles with momentum $p$ emitted from the freeze-out hypersurface. By boosting the latter to the particle rest frame (PRF) we obtain the information about the spin polarization which can be compared to the experimental data.
\medskip
%
\subsection{Phase-space density of the Pauli-Luba\'nski vector}
%
\begin{figure*}[t]
\centering
\includegraphics[width=8.8cm]{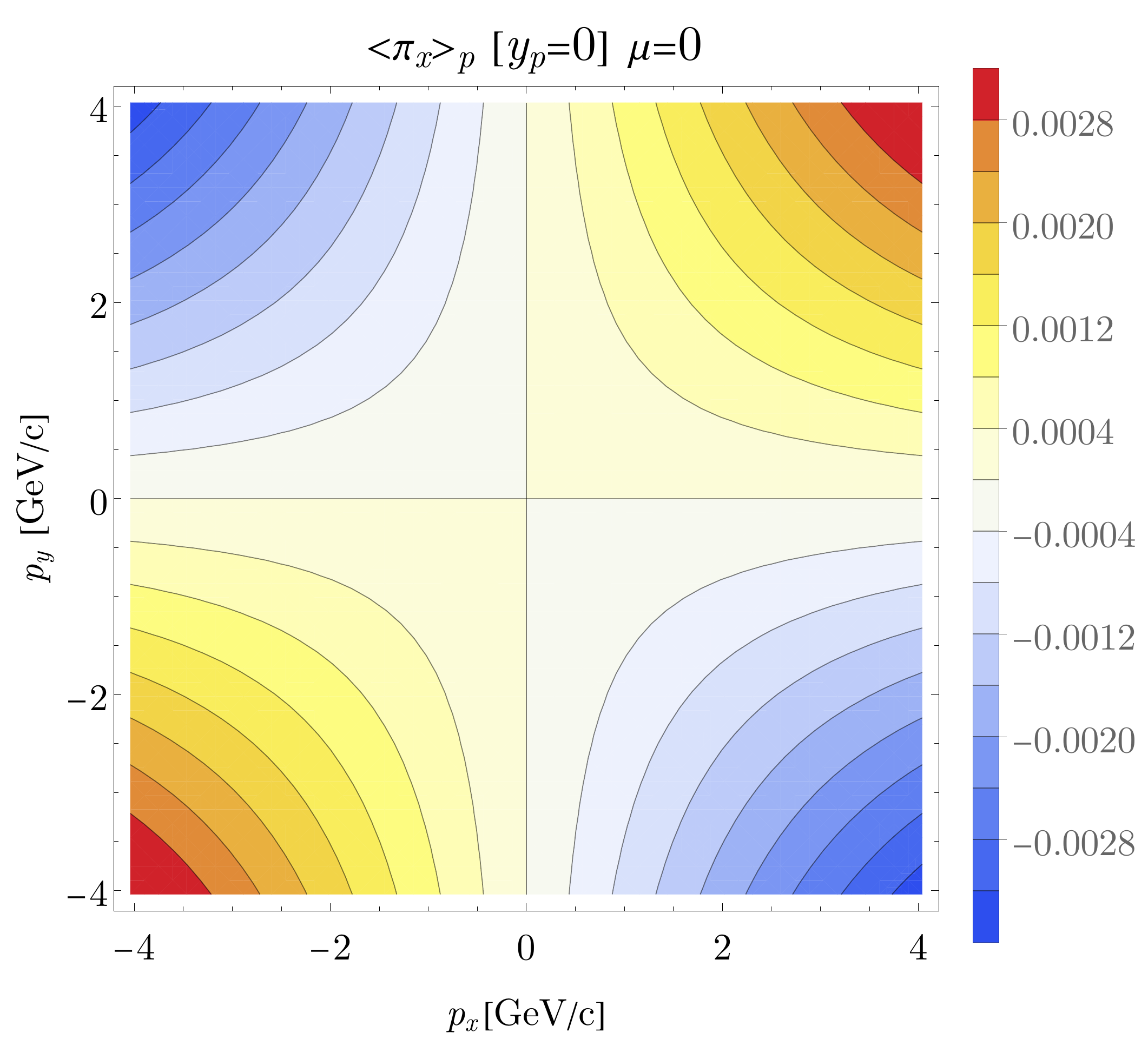}
\includegraphics[width=8.9cm]{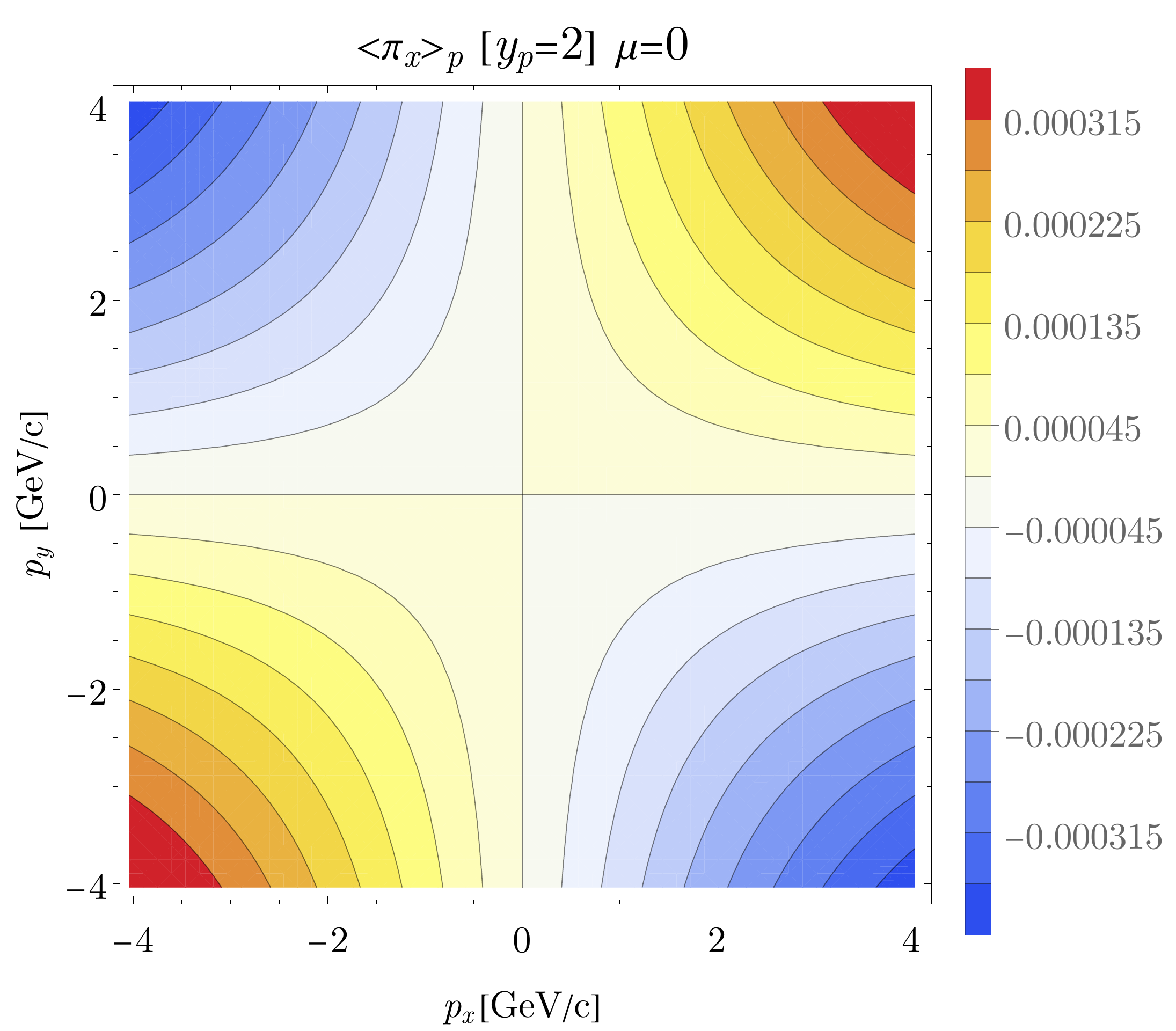}
\caption{Transverse momentum dependence of ${\langle\pi_{x}\rangle}$ for $y_p = 0$ (left) and for $y_p = 2$ (right).} 
\label{pixpol}
\end{figure*}
\begin{figure*}[t]
\centering
\includegraphics[width=8.9cm]{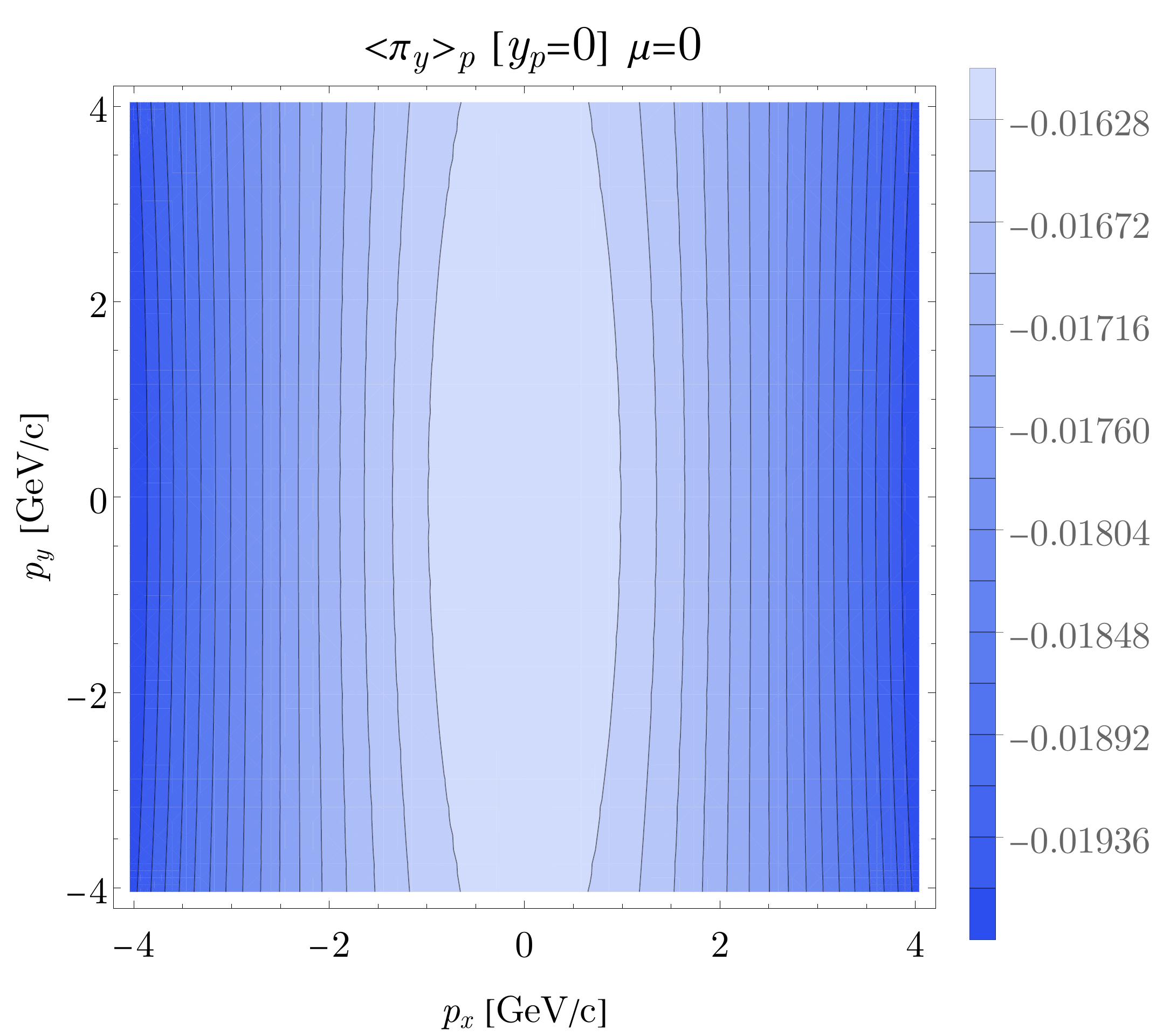}
\includegraphics[width=8.9cm]{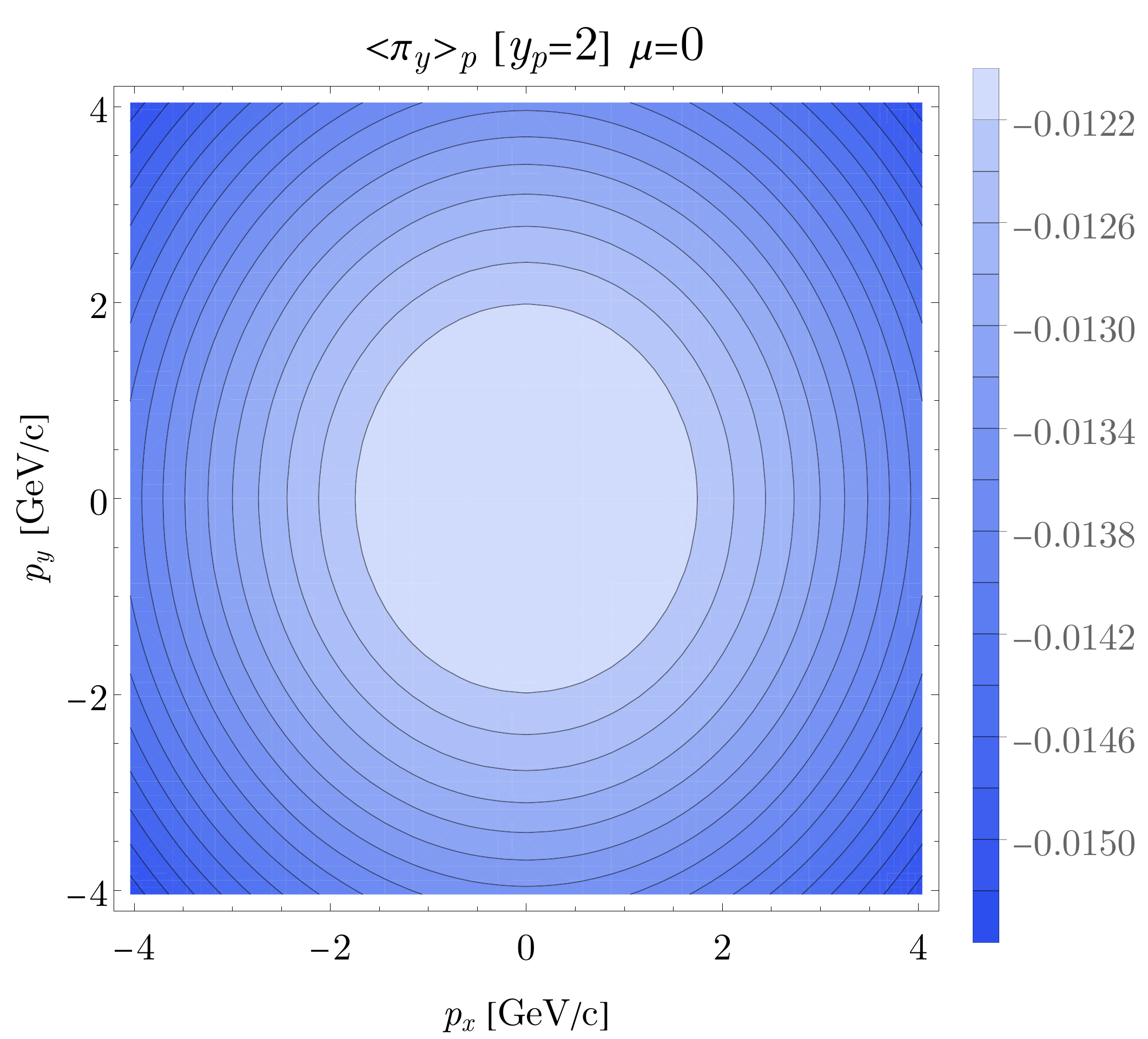}
\caption{Transverse momentum dependence of ${\langle\pi_{y}\rangle}$ for $y_p = 0$ (left) and for $y_p = 2$ (right).} 
\label{piypol}
\end{figure*}

The phase-space density of the PL four-vector $\Pi_{\mu}$ is given by the expression~\cite{Florkowski:2017dyn,Florkowski:2019qdp}
\begin{equation}
E_p\frac{d\Delta \Pi _{\mu }(x,p)}{d^3 p}
=-\frac{1}{2}\epsilon _{\mu \nu \alpha \beta }\Delta 
\Sigma _{\lambda }E_p\frac{dS^{\lambda ,\nu \alpha }(x,p)}
{d^3 p}\frac{p^{\beta }}{m},
\label{PL1}
\end{equation}
with $p^\beta$ being the four-momentum of the particle. The phase-space density of the spin tensor in \rf{PL1} in the GLW formulation is given by~\cite{Florkowski:2018ahw}
\beq
E_p \f{dS^{\lambda , \nu \a }}{d^3p} &=&\frac{{\cosh}(\xi)}{(2\pi)^3 } \, e^{-\beta \cdot p} p^{\lambda } \left(\omega ^{\nu\a}+\frac{2}{m^2} p^{[\nu }\omega ^{\a ]}{}_{\delta }p^{\delta } 
\right).
~~~~~~\label{eq:SGLW22}
\eeq
Substituting \rf{eq:SGLW22} into \rf{PL1} and integrating over the freeze-out hypersurface we obtain the total value of the momentum density of the PL four-vector
\begin{equation}
E_p\frac{d\Pi _{\mu }^{\mbox{*}}(p)}{d^3 p} = -\f{1}{(2 \pi )^3 m}
\int \cosh(\xi)
\Delta \Sigma _{\lambda } p^{\lambda } \,
e^{-\beta \cdot p} \,
(\widetilde{\omega }_{\mu \beta }p^{\beta})^{\mbox{*}},
\lab{PDLT}
\end{equation}
In the following we parameterize the four-momentum $p^\beta = \left(E_p, p_x, p_y, p_z \right)$ in terms of the transverse mass $m_T$, rapidity $y_p$, transverse momentum $p_T=\sqrt{p_x^2+p_y^2}$, and azimuthal angle $\phi_p$ as
\beq
E_p &=& m_T\ch(y_p)\, , \quad p_z = m_T\sh(y_p)\, ,\nn\\
p_x &=& p_T \cos({\phi_p})\, , \quad ~~p_y = p_T \sin({\phi_p})\,.
\label{pparameterize}
\eeq
As a result, one can write the argument of the thermal factor in \rf{PDLT} as
\begin{equation}
\beta  \cdot p = \frac{U \cdot p}{T} = \frac{m_T}{T}\cosh\left(y_p-\Phi \right),\lab{PU} 
\end{equation}
and the scalar product of particle four-momentum and element of freeze-out hypersurface at freeze-out proper time $\tau (\eta)$ as
\beq
\Delta \Sigma\cdot p 
&=& m_T \Big[ \tau(\eta) \cosh\left(y_p-\eta\right)\nn\\
&&~~~~~ -\tau^\prime(\eta)\sinh\left(y_p-\eta\right)\Big] dx \, dy \, d\eta.
\lab{SIGP}   
\eeq
Moreover, the contraction of the dual polarization tensor $\widetilde{\omega }_{\mu \beta}$ with $p^{\beta}$ in \rf{PDLT} gives rise to
\begin{widetext}
\begin{equation}
\widetilde{\omega }_{\mu \beta }p^{\beta }=\left[
\begin{array}{c}
\phantom{-}\left(C_{\kappa X} p_y-C_{\kappa Y} p_x\right)\sinh (\Phi)+\left(C_{\omega X} p_x+C_{\omega Y} p_y\right)\cosh (\Phi)+C_{\omega Z} m_T\sinh \left(y_p\right)\\ \\
-C_{\kappa Y} m_T \sinh \left(y_p-\Phi \right)-C_{\omega X} m_T \cosh \left(y_p-\Phi \right)+C_{\kappa Z} p_y \\ \\
C_{\kappa X} m_T\sinh \left(y_p-\Phi \right)-C_{\omega Y} m_T \cosh \left(y_p-\Phi \right)-C_{\kappa Z} p_x \\ \\
-\left(C_{\kappa X} p_y-C_{\kappa Y} p_x\right)\cosh(\Phi )-\left(C_{\omega X} p_x+C_{\omega Y} p_y\right)\sinh(\Phi)-C_{\omega Z} m_T\cosh\left(y_p\right) \\
\end{array}
\right]\,.
\lab{OP}
\end{equation}
\end{widetext}
Note that the spin polarization of $\Lambda$ hyperons measured in the experiments is defined in the rest frame of the decaying particle. Therefore, in order to compare our results with the experimental data, we have to Lorentz transform the quantity $\widetilde{\omega}_{\mu \beta} p^\beta$ to the PRF~\cite{Florkowski:2017dyn}. The result of the respective canonical boost~\cite{Leader:2001}, denoted in \rf{PDLT} by the asterisk, is
\begin{widetext}
\begin{equation}
(\tilde{\omega }_{\mu \beta }p^{\beta })^{\mbox{*}}=\left[
\begin{array}{c}
\phantom{-}0\\ \\
m~\alpha_p p_x p_y\left[C_{\kappa X} \sinh (\Phi )+C_{\omega Y} \cosh (\Phi )\right]\\ \\
m~\alpha_p p_y^2 \left[C_{\kappa X} \sinh (\Phi )+C_{\omega Y} \cosh (\Phi )\right]-m_T \left[C_{\kappa X} \sinh \left(\Phi -y_p\right)+C_{\omega Y} \cosh \left(\Phi -y_p\right)\right]\\ \\
-m~\alpha_p p_y \Big[m_T \Big(C_{\kappa X} \cosh \left(\Phi -y_p\right)+C_{\omega Y} \sinh \left(\Phi -y_p\right)\Big)+m \Big(C_{\kappa X} \cosh (\Phi )+C_{\omega Y} \sinh (\Phi )\Big)\Big] \\
\end{array}
\right]\,,
\lab{OP3}
\end{equation}
\end{widetext}
where we defined $\alpha_p \equiv 1/(m^2 + m E_p)$~\cite{Florkowski:2017dyn}. Above we have neglected all spin coefficients except for $C_{\kappa X}$ and $C_{\omega Y}$ due to the fact that for the non-boost-invariant and transversely homogeneous systems only these are relevant for the physical scenario under study (for general form of $\widetilde{\omega}_{\mu \beta} p^\beta$ in PRF for the non-boost invariant and transversely homogeneous system see App.~\ref{app:pol}).
%
\subsection{Average polarization per particle}
%
%
\begin{figure}[t]
\begin{center}
\includegraphics[width=8.4cm]{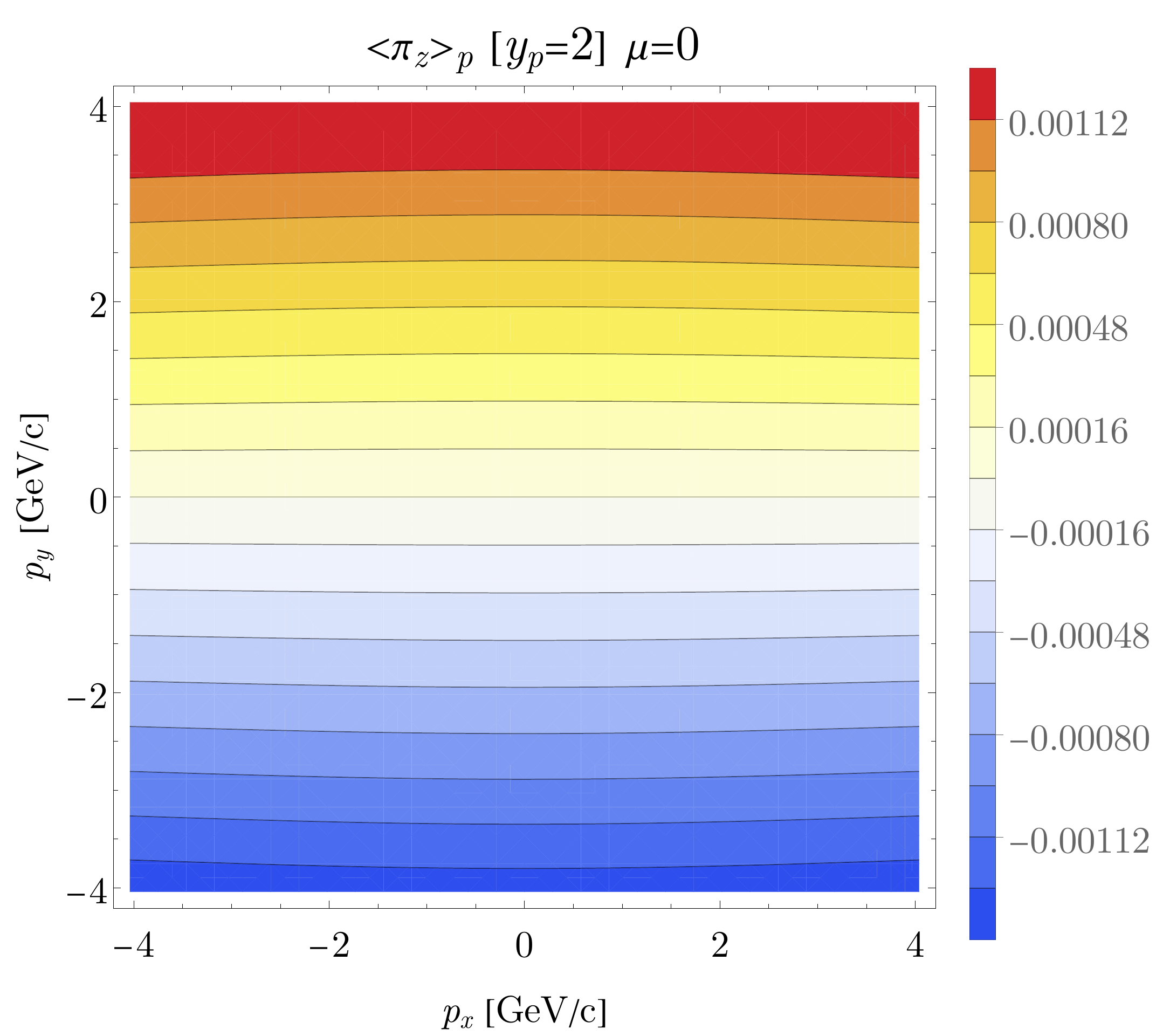}
\end{center}
\caption{Transverse momentum dependence of ${\langle\pi_{z}\rangle}$ for  $y_p = 2$.}
\label{pizpol}
\end{figure}
The  mean spin polarization per particle $\langle\pi_{\mu}\rangle_p$  is defined as the ratio of the momentum density of the total PL four-vector \rfn{PDLT} over the momentum density of particles~\cite{Florkowski:2018ahw} 
\beq
{\langle\pi_{\mu}\rangle}_p=\frac{E_p\frac{d\Pi _{\mu }^{\mbox{*}}(p)}{d^3 p}}{E_p\frac{d{\cal{N}}(p)}{d^3 p}},
\label{meanspin}
\eeq
where the quantity in the denominator is given by the formula
\beq
E_p\frac{d{\cal{N}}(p)}{d^3 p}&=&
\f{4}{(2 \pi )^3}
\int \Delta  \Sigma _{\lambda } p^{\lambda } \cosh(\xi)
\,
e^{-\beta \cdot p} \,.
\label{MD}
\eeq
%
\subsection{Momentum dependence of the polarization}
%
In Figs.~\ref{pixpol}-\ref{pizpol} we show the results of the components of the mean spin polarization vector calculated using \rf{meanspin} at mid ($y_p = 0$) and forward ($y_p = 2$) rapidities assuming zero baryon chemical potential in the background. 
\smallskip

From Fig.~\ref{pixpol}, we see that ${\langle\pi_{x}\rangle}_p$ component has always quadrupole structure with its sign changing sequentially through the quadrants, and the magnitude decreasing with rapidity. The $p_x p_y$ structure seen in Fig.~\ref{pixpol} can be inferred also from  $x$ component of $(\tilde{\omega }_{\mu \beta }p^{\beta })^{\mbox{*}}$   \rfn{OP3}, and it was observed already in the case of Bjorken-flowing background \cite{Florkowski:2018fap,Florkowski:2019qdp}.
\smallskip

In Fig. \ref{piypol} we present ${\langle\pi_{y}\rangle}_p$ which, in agreement with the chosen initialization of the model, is negative indicating the direction of the spin angular momentum three-vector opposite to the standard orientation of $y$ axis in heavy-ion collisions. At forward rapidities the values of ${\langle\pi_{y}\rangle}_p$ slowly decrease and become almost $\phi_p$ independent.

A particularly interesting observable which is measured in experiments and still awaits theoretical explanation, is the longitudinal polarization, that is the polarization along the beam direction $\langle\pi_{z}\rangle_p$~\cite{STAR:2019erd,ALICE:2021pzu}. Some information about the longitudinal spin polarization can be obtained already from the symmetry considerations of the integral in \rf{PDLT}. One can easily notice that due to the symmetric integration region in $\eta$ only $\eta$-even integrands will give non-vanishing contribution. In the case of $z$ component of $\langle\pi_{\mu}\rangle_p$ at $y_p = 0$, the assumption of $C_{\kappa X}$ being odd and $C_{\omega Y}$ being even function of $\eta$ leads to the conclusion that the integral $\langle\pi_{z}\rangle_p$ at mid rapidity in \rf{PDLT} will give zero, see also \rf{OP3}. Interestingly, when treated differentially, at forward rapidities we find a non-trivial longitudinal polarization pattern, see Fig. \ref{pizpol}. Due to our initialization, after integrating over transverse momenta the latter, as well as in the case of $x$ component, gives zero.

Obviously, the numerical results presented here do not reproduce the quadrupole structure of the longitudinal polarization as seen in experiment which is largely due to assumed homogeneity in the transverse plane. It is possible that the quadrupole structure of longitudinal spin polarization $\langle\pi_{z}\rangle_p$ at mid rapidity~\cite{STAR:2019erd} will be present in full (3+1)D geometry due to the presence of elliptic flow resulting from the elliptic deformation of the system in the transverse plane~\cite{Voloshin:2017kqp}. In the spin-thermal models the quadrupole structure in spin polarization (with opposite sign in comparison to experiments) arises due to polarization-vorticity coupling~\cite{Karpenko:2016jyx}.

Since the effect of homogeneous $\mu$ profile, as shown in Fig.~\ref{CkxCwy}, is small, we observe no qualitative difference in the momentum dependence of mean spin polarization of particles as well. Hence, we abstain from presenting the respective plots of $\langle\pi_{\mu}\rangle_p$ in here.

%
\subsection{Momentum averaged polarization}
\label{sec:GP}
%
Here we turn to discussion of momentum-averaged polarization which, using \rf{meanspin}, may be expressed as
\beq
\langle\pi_{\mu}\rangle = \frac{\int dP {\langle\pi_{\mu}\rangle}_p E_p\frac{d{\cal{N}}(p)}{d^3 p}}{\int dP E_p\frac{d{\cal{N}}(p)}{d^3 p}} \equiv \frac{\int d^3 p ~\frac{d\Pi_{\mu }^{\mbox{*}}(p)}{d^3 p}}{\int d^3 p~\frac{d{\cal{N}}(p)}{d^3 p}}.
\label{AvgPol}
\eeq
Due to our choice of initialization and symmetry properties of spin coefficients with respect to $\eta$, we find from Eq.~\rfn{AvgPol}, that only $\langle \pi_y \rangle$ component gives non-zero result after performing integration over freeze-out hypersurface and transverse momentum coordinates~\cite{STAR:2017ckg,Adam:2018ivw,STAR:2019erd,Acharya:2019ryw,Kornas:2019}.
%
\begin{figure*}[t]
\centering
\includegraphics[width=8.7cm]{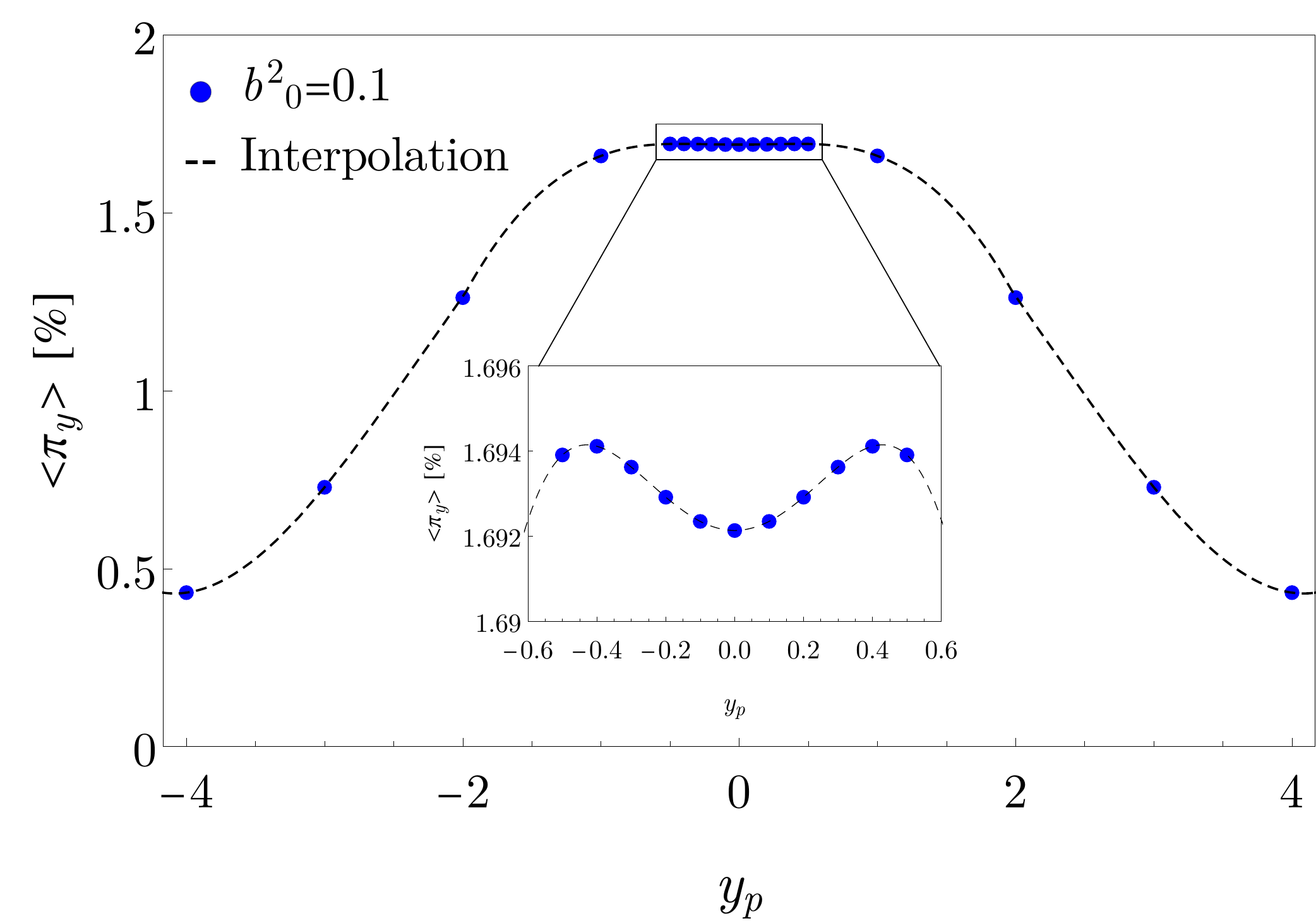}
\includegraphics[width=8.7cm]{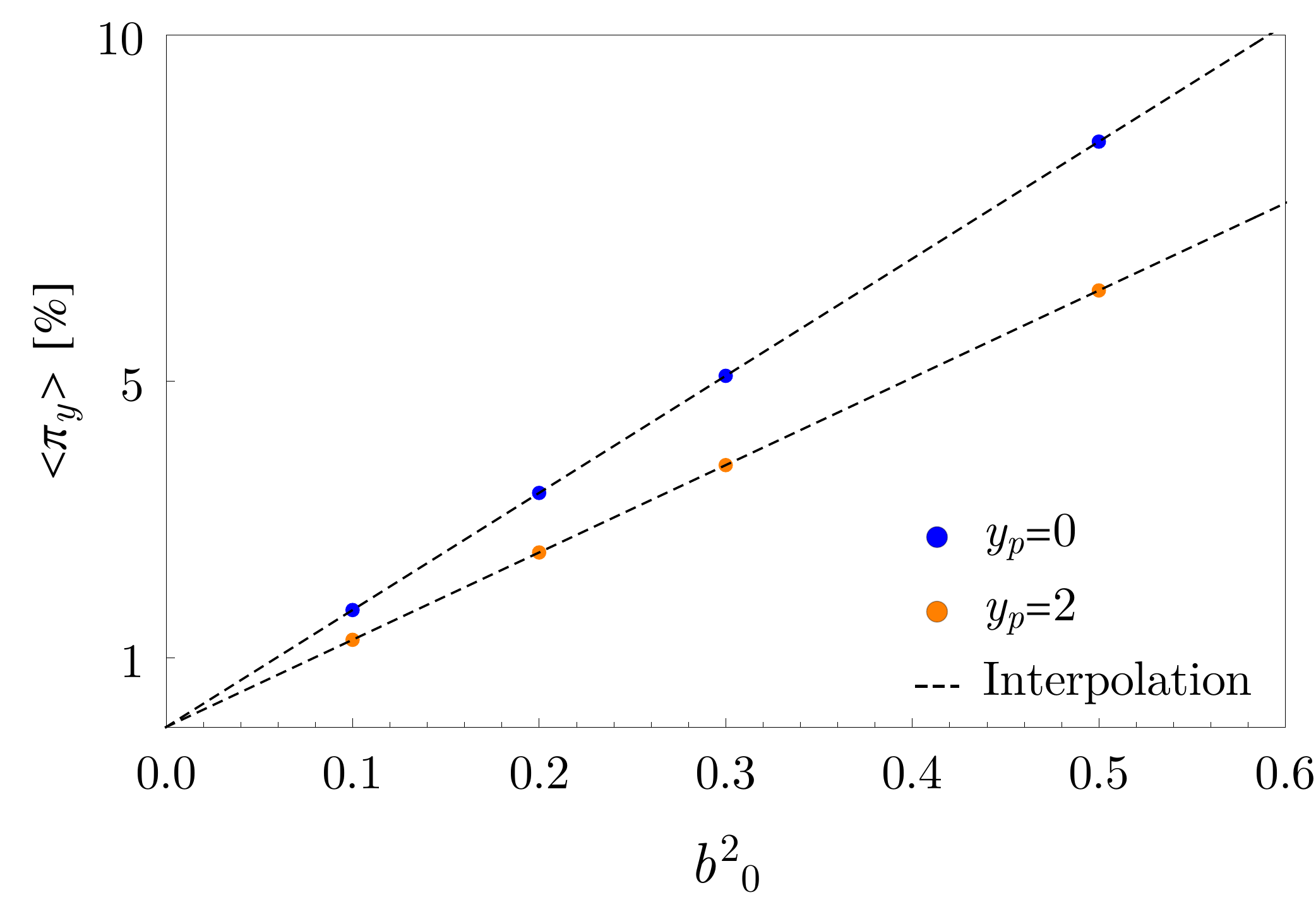}
\caption{$\langle \pi_y \rangle$ component of momentum averaged polarization as a function of rapidity (left) and as a function of spin coefficient
$b^2_{0}$
(right).}
\label{GlobalPol}
\end{figure*}
\begin{figure*}[t]
\centering
\includegraphics[width=8.9cm]{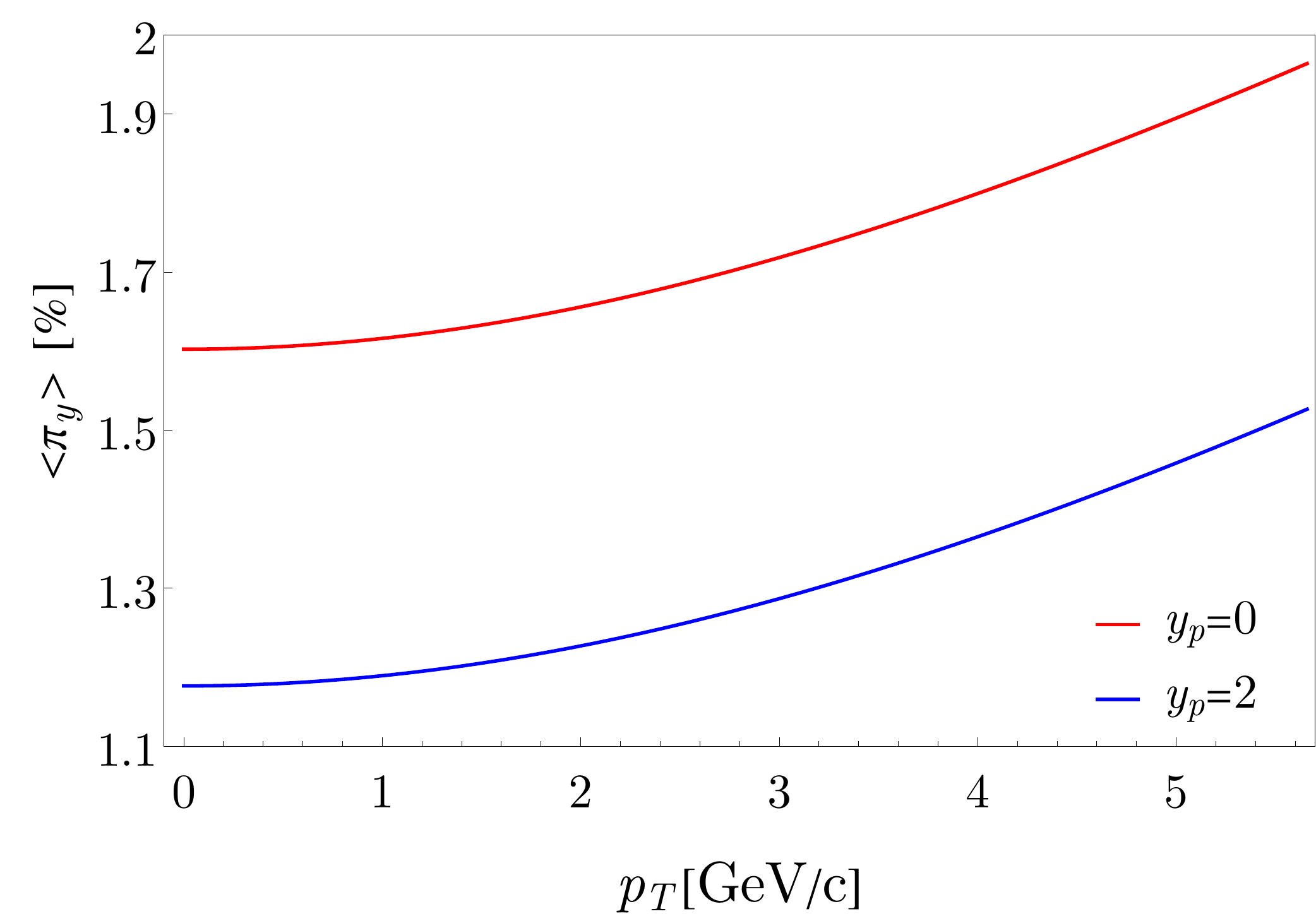}
\includegraphics[width=8.9cm]{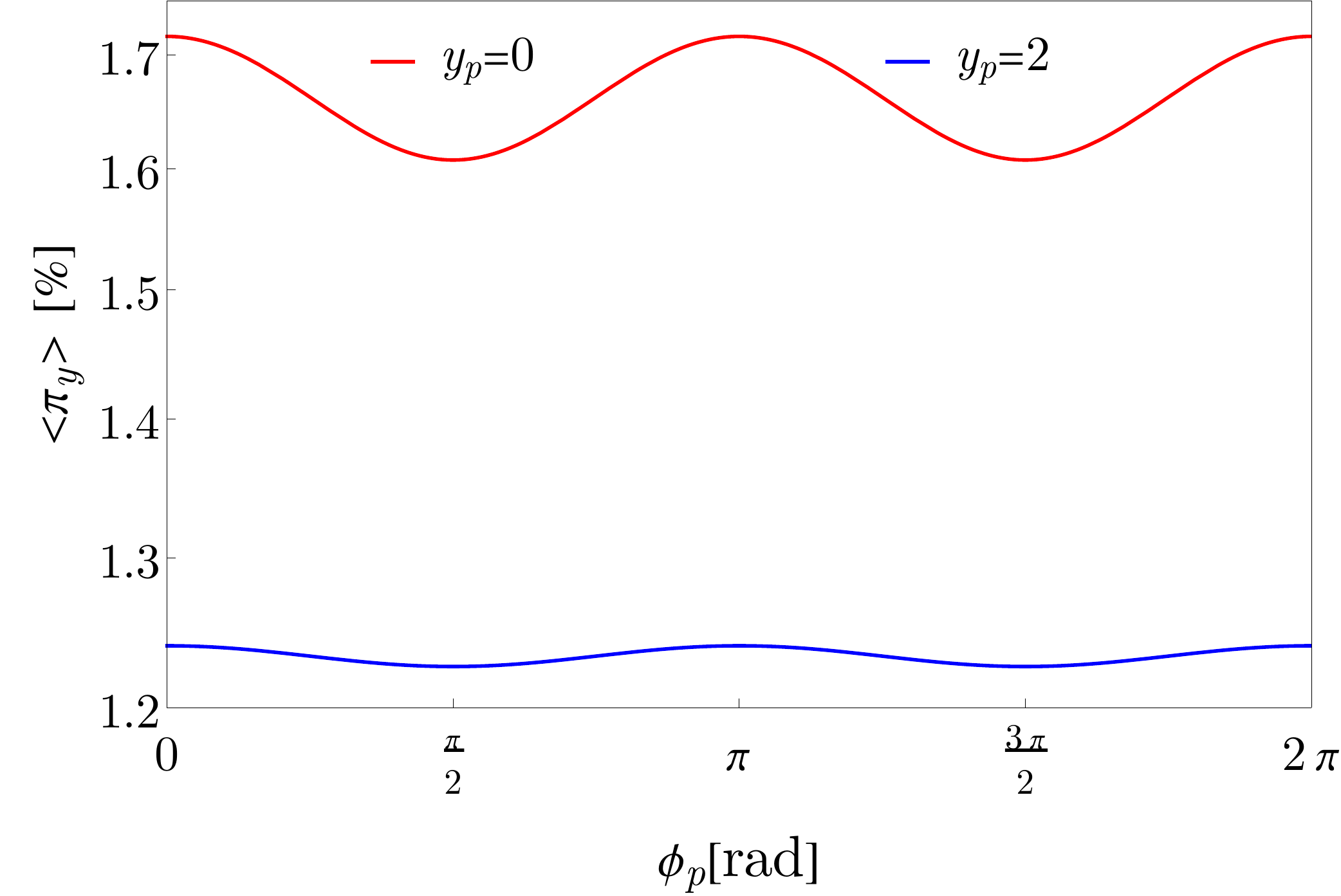}
\caption{$p_T$ (left) and $\phi_p$ 
(right)
dependent $\langle \pi_y \rangle$ component of the momentum averaged polarization.}
\label{pTpol}
\end{figure*}

In Fig.~\ref{GlobalPol} (left panel) we present the behavior of the transverse momentum integrated $y$ component of mean polarization as a function of rapidity. One can observe that in the mid-rapidity region the magnitude of $\langle \pi_y \rangle$ component is around 1.6\% and then decreases with increasing rapidity. This qualitative behavior is also observed in other models~\cite{Wei:2018zfb} and will be subject to future STAR measurements~\cite{xu:2021imi}. In Fig.~\ref{GlobalPol} (right panel) we show the relation between $\langle \pi_y \rangle$ and the  initial central value of $b^2$ component calculated for two rapidity values $y_p = 0$ and $y_p = 2$.
In Fig.~\ref{pTpol}, we show transverse momentum $p_T$ and azimuthal angle $\phi_p$ dependence of the double differential $\langle \pi_y \rangle$ spin polarization component. We find the polarization dependence on $p_T$ 
rather strong in comparison to experiments~\cite{STAR:2018gyt} and other models~\cite{Wei:2018zfb}, see Fig. \ref{pTpol} (left panel). The behavior of the momentum averaged polarization as a function of $\phi_p$ is depicted in Fig. \ref{pTpol} (right panel). 

Figs.~\ref{withMuplots1} show the $\langle \pi_y \rangle$ component of momentum averaged polarization as a function of $y_p$, $p_T$, and $\phi_P$
with non-zero baryon chemical potential evolution. Similarly as in the case of hydrodynamic variables see small effects on the spin polarization vector.
\section{Summary}
\label{sec:summ}
In this paper we have analyzed the space-time evolution of spin polarization for non-boost-invariant and transversely homogeneous system based on the framework of spin hydrodynamics~\cite{Florkowski:2017ruc,Florkowski:2017dyn, Florkowski:2018ahw}. We have found that, in contrast to Ref.~\cite{Florkowski:2019qdp} where we had studied the spin polarization evolution undergoing Bjorken expansion, in the current analysis some of the spin coefficients affect each other's behavior. In the current setup, we have used relativistic ideal gas equation of state for $\Lambda$ particles. 

We have calculated the momentum dependent and momentum averaged components of mean spin polarization vector for $\Lambda$ particles at mid ($y_p = 0$) and forward ($y_p = 2$) 
rapidities, and we have found, as expected, that only $y$-component of the momentum averaged spin polarization is non-zero. We have also shown that the $p_T$ and $\phi_p$
dependence of spin polarization exhibits some non trivial features. In particular, we have observed an interesting decay of the $\langle\pi_{y}\rangle$ at forward rapidities.

The analysis done in this paper indicates that correct description of the measured observables may require breaking of the symmetry in the transverse plane and performing modelling in the full (3+1)-dimensional setup. The studies along these lines are left for future analysis.
%
\begin{acknowledgments}
This research was supported in part by the Polish National Science Center Grants No. 2016/23/B/ST2/00717 and No. 2018/30/E/ST2/00432. We would like to thank the organizers of the online Workshop on QGP Phenomenology by Institute for Research in Fundamental Sciences, Iran during which the discussions with participants helped us to gain some new perspective.
\end{acknowledgments}

\appendix
\begin{widetext}
\section{Derivation of evolution equations for spin coefficients for an arbitrary (3+1)-dimensional system}
\label{app:deriv}
%
\subsection{Spin tensor decomposition}
%
In the derivation of the equations of motion for the spin coefficients for an arbitrary (3+1)-dimensional [(3+1)D] geometry it is convenient to use the  decomposition of the spin tensor \rf{SGLW2} in the four-vector basis. 
Subsequently we decompose the spin tensor \rf{SGLW2} in the basis $U_\a$, $X_\a$, $Y_\a$, and $Z_a$ by projecting on all possible combinations of these four-vectors. In this way one obtains
\beq
S^{\a,\b\g}=\sum_{i=x,y,z} S_{\a_i}^{\a,\b\g}+ S_{\b_i}^{\a,\b\g}
 \label{Sdiv}
\eeq
where
\beq
S_{\a_x}^{\a,\b\g}&=&2~\a_{x1} 
~U^\a ~U^{[\b} X^{\g]} + 
\a_{x2} \LS Y^\a ~Y^{[\b} X^{\g]} + Z^\a ~Z^{[\b} X^{\g]}
\RS, \label{AX}\\
S_{\a_y}^{\a,\b\g}&=&2~\a_{y1} 
~U^\a ~U^{[\b} Y^{\g]} +
\a_{y2} \LS X^\a ~X^{[\b} Y^{\g]} + Z^\a ~Z^{[\b} Y^{\g]} \RS, \label{AY}\\
S_{\a_z}^{\a,\b\g}&=&2~\a_{z1}
~U^\a ~U^{[\b} Z^{\g]} +
\a_{z2} \LS X^\a ~X^{[\b} Z^{\g]} + Y^\a ~Y^{[\b} Z^{\g]} \RS,
 \label{AZ}\\
S_{\b_x}^{\a,\b\g}&=&2~\b_{x1} \Big[
Y^\a ~U^{[\b} Z^{\g]} +
Z^\a ~Y^{[\b} U^{\g]}
\Big] - 
2~\b_{x2} U^\a ~Y^{[\b} Z^{\g]}\, ,
 \label{BX}\\
S_{\b_y}^{\a,\b\g}&=&2~\b_{y1} \Big[ 
Z^\a ~U^{[\b} X^{\g]} +  
X^\a ~Z^{[\b} U^{\g]} 
\Big] - 
2~\b_{y2} U^\a ~Z^{[\b} X^{\g]}\, ,
 \label{BY}\\
S_{\b_z}^{\a,\b\g}&=&2~\b_{z1} \Big[
X^\a ~U^{[\b} Y^{\g]} +
Y^\a ~X^{[\b} U^{\g]}
\Big] - 
2~\b_{z2} U^\a ~X^{[\b} Y^{\g]} \, .
 \label{BZ}
\eeq
and $\a_i$ and $\b_i$ are defined through the Eqs. \rfn{alpha-i}-\rfn{beta-i2}.
%
\medskip
\subsection{Divergence of the spin tensor }
%
Calculating the partial derivative of Eqs.~\rfn{AX}-\rfn{BZ} we obtain the following six expressions (see the notation in Sec.~\ref{sec:convention})
\beq
\p_\a S_{\a_x}^{\a,\b\g} &=& 
2~\UD{\a_{x1}} ~U^{[\b} X^{\g]} +
\YD{\a_{x2}} ~Y^{[\b} X^{\g]} + 
\ZD{\a_{x2}} ~Z^{[\b} X^{\g]} 
+ 2~\a_{x1} \LS
\tU ~U^{[\b} X^{\g]} + \UD{U^{[\b}} X^{\g]} + 
  U^{[\b} \UD{X^{\g]}} \RS \nn\\
&+&
\a_{x2} \LS \tY ~Y^{[\b} X^{\g]} + 
\tZ ~Z^{[\b} X^{\g]} +
\YD{Y^{[\b}} X^{\g]} +
  Y^{[\b} \YD{X^{\g]}} +
  \ZD{Z^{[\b}} X^{\g]} +
Z^{[\b} \ZD{X^{\g]}}
\RS\,,
\label{DAX}\\
\p_\a S_{\a_y}^{\a,\b\g} &=& 
2~\UD{\a_{y1}}~U^{[\b} Y^{\g]} + 
\XD{\a_{y2}}~X^{[\b} Y^{\g]} + 
\ZD{\a_{y2}}~Z^{[\b} Y^{\g]} 
+ 2~\a_{y1} \LS
\tU ~U^{[\b} Y^{\g]} +
\UD{U^{[\b}} Y^{\g]} + U^{[\b} \UD{Y^{\g]}} \RS
\nn \\
 &+&
\a_{y2} \LS \XD{X^{[\b}} Y^{\g]} + X^{[\b} \XD{Y^{\g]}} + 
\ZD{Z^{[\b}} Y^{\g]}  + Z^{[\b} \ZD{Y^{\g]}} + \tX ~X^{[\b} Y^{\g]} +
\tZ ~Z^{[\b} Y^{\g]}
\RS \,,
\label{DAY}\\
\p_\a S_{\a_z}^{\a,\b\g} &=& 
2~\UD{\a_{z1}} ~U^{[\b} Z^{\g]} + 
\XD{\a_{z2}} ~X^{[\b} Z^{\g]} + 
\YD{\a_{z2}} ~Y^{[\b} Z^{\g]} 
+2~\a_{z1} \LS
\tU ~U^{[\b} Z^{\g]} + \UD{U^{[\b}} Z^{\g]} +  U^{[\b} \UD{Z^{\g]}} \RS
\nn \\
&+&
\a_{z2} \LS \XD{X^{[\b}} Z^{\g]} +  X^{[\b} \XD{Z^{\g]}}  + 
\YD{Y^{[\b}} Z^{\g]} +Y^{[\b} \YD{Z^{\g]}}
+ \tX ~X^{[\b} Z^{\g]} + 
\tY~Y^{[\b} Z^{\g]} 
\RS\,,
\label{DAZ}
\eeq
\beq
\p_\a S_{\b_x}^{\a,\b\g} &=& 2\LS 
\YD{\b_{x1}} ~U^{[\b} Z^{\g]} +  
\ZD{\b_{x1}} ~Y^{[\b} U^{\g]}
+ \b_{x1} \LR
 \tY ~U^{[\b} Z^{\g]} +  
 \tZ ~Y^{[\b} U^{\g]} 
\RR
- \UD{\b_{x2}} ~Y^{[\b} Z^{\g]} -
\b_{x2} \tU ~Y^{[\b} Z^{\g]}
\RS
\nn \\
 &+&  2~\b_{x1} \LS 
 \YD{U^{[\b}} Z^{\g]} +  
 \ZD{Y^{[\b}} U^{\g]} + U^{[\b} \YD{Z^{\g]}} +  
   Y^{[\b} \ZD{U^{\g]}} \RS - 
2~\b_{x2} \LR \UD{Y^{[\b}} Z^{\g]} + Y^{[\b} \UD{Z^{\g]}} \RR,
\label{DBX}\\
\p_\a S_{\b_y}^{\a,\b\g} &=& 2\LS 
\ZD{\b_{y1}} ~U^{[\b} X^{\g]} +  
\XD{\b_{y1}} ~Z^{[\b} U^{\g]} +
\b_{y1} \LR 
 \tZ ~U^{[\b} X^{\g]} +  
 \tX ~Z^{[\b} U^{\g]} 
\RR 
- \UD{\b_{y2}} ~Z^{[\b} X^{\g]} - \b_{y2} \tU ~Z^{[\b} X^{\g]}
\RS 
\nn \\
 &+&  2~\b_{y1} \LS 
  \ZD{U^{[\b}} X^{\g]} +  
  \XD{Z^{[\b}} U^{\g]} +  U^{[\b} \ZD{X^{\g]}} +  
  Z^{[\b} \XD{U^{\g]}}
\RS - 
2~\b_{y2} \LR \UD{Z^{[\b}} X^{\g]} + Z^{[\b} \UD{X^{\g]}} \RR \, ,
\label{DBY}\\
\p_\a S_{\b_z}^{\a,\b\g} &=& 2\LS 
\XD{\b_{z1}} ~U^{[\b} Y^{\g]} +
\YD{\b_{z1}} ~X^{[\b} U^{\g]} + \b_{z1} \LR 
 \tX ~U^{[\b} Y^{\g]} +  
 \tY ~X^{[\b} U^{\g]}
\RR
-\UD{\b_{z2}} ~X^{[\b} Y^{\g]} - \b_{z2} \tU ~X^{[\b} Y^{\g]}\RS
\nn \\
 &+&  2~\b_{z1} \LS 
  \XD{U^{[\b}} Y^{\g} +  
  \YD{X^{[\b}} U^{\g]} + U^{[\b} \XD{Y^{\g]}} +  
  X^{[\b} \YD{U^{\g]}} 
\RS - 
2~\b_{z2} \LR \UD{X^{[\b}} Y^{\g]} + X^{[\b} \UD{Y^{\g]}} \RR\,.
\label{DBZ}
\eeq
which, after coupling together, yield
\beq
\p_\a S_{\rm GLW}^{\a,\b\g} = \p_\a S_{\a_x}^{\a,\b\g} + \p_\a S_{\a_y}^{\a,\b\g} + \p_\a S_{\a_z}^{\a,\b\g} + \p_\a S_{\b_x}^{\a,\b\g} + \p_\a S_{\b_y}^{\a,\b\g} + \p_\a S_{\b_z}^{\a,\b\g} = 0.
\label{DSGLW}
\eeq
Note that, all the definitions of derivatives and divergences are defined through the relations \rf{deriv} and \rf{div}. In order to obtain the final evolution equations for the spin coefficients we contract the \rf{DSGLW} with $U_\b X_\g$, $U_\b Y_\g$, $U_\b Z_\g$,  $Y_\b Z_\g$, $X_\b Z_\g$ and $X_\b Y_\g$ and obtain the following expressions
\beq
&&-\UD{\a_{x1}} -\a_{x1} \tU -\frac{\a_{x2}}{2} \LR  U\YD{Y} + U\ZD{Z} \RR
+ \frac{\a_{y2}}{2} U\XD{Y} + \a_{y1} X\UD{Y}  
+ \frac{\a_{z2}}{2} U\XD{Z} + \a_{z1} X\UD{Z} + \b_{x1} \LR X\YD{Z} - X\ZD{Y} \RR - \ZD{\b_{y1}} - \b_{y1} \LR \tZ + X\XD{Z} \RR
\nn\\
&& 
 + \b_{y2}  U\UD{Z} 
+ \YD{\b_{z1}} + \b_{z1} \LR \tY + X\XD{Y} \RR - \b_{z2}  U\UD{Y}   =0, \label{UX}\\
&&-\UD{\a_{y1}} -\a_{y1} \tU -\frac{\a_{y2}}{2} \LR U\XD{X} + U\ZD{Z} \RR 
+\frac{\a_{z2}}{2} U\YD{Z}
+ \a_{z1} Y\UD{Z} 
+\frac{\a_{x2}}{2} U\YD{X} + \a_{x1} Y\UD{X} + \b_{y1} \LR Y\ZD{X} - Y\XD{Z} \RR 
- \XD{\b_{z1}} - \b_{z1} \LR \tX + Y\YD{X} \RR \nn\\
&&
+ \b_{z2}  U\UD{X} 
+ \ZD{\b_{x1}} + \b_{x1} \LR \tZ + Y\YD{Z} \RR - \b_{x2}  U\UD{Z}  =0, \label{UY} \\
&&-\UD{\a_{z1}} -\a_{z1} \tU -\frac{\a_{z2}}{2} \LR U\XD{X} + U\YD{Y} \RR 
+ \frac{\a_{x2}}{2} U\ZD{X} + \a_{x1} Z\UD{X}
+ \frac{\a_{y2}}{2} U\ZD{Y} + \a_{y1} Z\UD{Y}+ \b_{z1} \LR Z\XD{Y} - Z\YD{X} \RR 
- \YD{\b_{x1}} 
- \b_{x1} \LR \tY + Z\ZD{Y} \RR  \nn\\
&&
+ \b_{x2}  U\UD{Y} 
+ \XD{\b_{y1}} 
+ \b_{y1} \LR \tX +  Z\ZD{X} \RR 
- \b_{y2}  U\UD{X}   =0,  \label{UZ}\\
&&-\frac{\ZD{\a_{y2}}}{2} - \frac{\a_{y2}}{2} \LR \tZ - Z\XD{X} \RR + \a_{y1} Z\UD{U} 
+\frac{\YD{\a_{z2}}}{2} + \frac{\a_{z2}}{2} \LR \tY - Y\XD{X} \RR - \a_{z1} Y\UD{U}
+ \frac{\a_{x2}}{2} \LR Y\ZD{X}- Z\YD{X} \RR - \UD{\b_{x2}} -\b_{x2} \tU \nn  \\
&& - \b_{x1} \LR Y\YD{U} + Z\ZD{U} \RR 
+\b_{y1}   Y\XD{U} - \b_{y2} Y\UD{X} 
+\b_{z1}   Z\XD{U} - \b_{z2} Z\UD{X} =0, \label{YZ}\\
&&-\frac{\XD{\a_{z2}}}{2} - \frac{\a_{z2}}{2} \LR \tX - X\YD{Y} \RR + \a_{z1} X\UD{U} 
+ \frac{\ZD{\a_{x2}}}{2} + \frac{\a_{x2}}{2} \LR \tZ - Z\YD{Y} \RR - \a_{x1} Z\UD{U}
+ \frac{\a_{y2}}{2} \LR Z\XD{Y}- X\ZD{Y} \RR - \UD{\b_{y2}} -\b_{y2} \tU
\nn   \\
&&- \b_{y1} \LR Z\ZD{U} + X \XD{U}\RR 
+\b_{z1}   Z\YD{U} - \b_{z2} Z\UD{Y} 
+\b_{x1}   X\YD{U} - \b_{x2} X\UD{Y} =0, \label{ZX}\\
&&- \frac{\YD{\a_{x2}}}{2} - \frac{\a_{x2}}{2} \LR \tY - Y\ZD{Z} \RR + \a_{x1} Y\UD{U} 
+ \frac{\XD{\a_{y2}}}{2} + \frac{\a_{y2}}{2} \LR \tX - X\ZD{Z} \RR - \a_{y1} X\UD{U} 
+ \frac{\a_{z2}}{2} \LR X\YD{Z}- Y\XD{Z} \RR - \UD{\b_{z2}} -\b_{z2} \tU
\nn \\
&& - \b_{z1} \LR X\XD{U} + Y\YD{U} \RR 
+\b_{x1}   X\ZD{U} - \b_{x2} X\UD{Z} 
+\b_{y1}   Y\ZD{U} - \b_{y2} Y\UD{Z} =0\,.
 \label{XY}
\eeq
respectively. One should stress that equations of motion \rfmtwo{UX}{XY} are valid for an arbitrary 3+1D system with no symmetries imposed.
For our non-boost invariant and transversely homogeneous system these equations get simplified  to \rfmtwo{spineq1}{spineq6}. One can also check that in the case of Bjorken expanding system the spin evolution equations \rfmtwo{spineq1}{spineq6} further simplify to~\cite{Florkowski:2019qdp}
\beq
&&\UD{\a_{x1}} = -\a_{x1} \tU -\frac{\a_{x2}}{2} U\ZD{Z}, \nn\\
&&\UD{\a_{y1}} =-\a_{y1} \tU -\frac{\a_{y2}}{2} U\ZD{Z}, \nn\\
&&{\UD{\a_{z1}} = -\a_{z1} \tU}, \nn\\
&&{\UD{\b_{x2}} = -\b_{x2} \tU} - \b_{x1} ~{Z\ZD{U}}, \nn \\
&&{\UD{\b_{y2}}= -\b_{y2} \tU} - \b_{y1}~ {Z\ZD{U}},\nn\\
&&{\UD{\b_{z2}} = -\b_{z2} \tU},
\label{BJEq}
\eeq
where you can notice that each spin coefficient evolves independently of others.
%
\section{Physical interpretation of the spin coefficients}
\label{app:spincoef}
%
This section discusses the contribution from the orbital and spin angular momentum to the total angular momentum at the surface of constant longitudinal proper time $\tau={\tau}_{\rm FO}$. To get some physical insights about the interpretation of the spin coefficients $\boldsymbol{C_{\kappa}}$ and $\boldsymbol{C_{\omega}}$, we calculate the spin angular momentum components at the freeze-out. The following calculations will also help us to understand which spin coefficients are important according to the physics assumed in our current set-up for the evolution of the spin polarization.
%
\subsection{Orbital contribution}
\label{app:orbitalcontri}
%
We consider a region of spacetime defined by the constant longitudinal proper time ${\tau}_{\rm FO}$ with  $-\eta_{\rm FO}/2 \leq \eta \leq +\eta_{\rm FO}/2$ and $\sqrt{x^2+y^2} \leq R_{\rm FO}$.

The orbital angular momentum  is given by the following expression
\beq
L^{\mu\nu}_{\rm FO} =\int
d \Sigma _{\lambda } L^{\lambda,\mu\nu} =\int
d \Sigma _{\lambda } \lt(x^{\mu}T^{\lambda\nu}
-x^{\nu}T^{\lambda\mu} \rt),
\label{OAM}
\eeq
where
\beq
d \Sigma _{\lambda } &=& \tau_{\rm FO}\, U_{\lambda}^{\rm B}\, dx dy\, d\eta \lab{sig} 
\eeq
is the element of the hypersurface of constant proper time $\tau_{\rm FO}$ with $U_{\lambda}^{\rm B} = \big(\!\cosh \eta, 0,0, \sinh \eta\big)$.
Using Eq.~\rfn{Tmn} we can write,
\beq
L^{\mu\nu}_{\rm FO} &=&A_\perp \tau_{\rm FO} \int d\eta\,  U_{\lambda}^B U^{\lambda}  \,\Bigg[\Big({\cal E} + {\cal P}\Big) \left(x^{\mu}U^{\nu}-x^{\nu}U^{\mu}\right) -{\cal P} \left(x^{\mu}U^{\nu}_{\rm B} -x^{\nu}U^{\mu}_{\rm B} \right)\Bigg],
\eeq
where $A_\perp=\pi R^2_{\rm FO}$. Using the form of $U^{\mu}$ from \rf{flow} and transverse homogeneity of the system we can show that
\beq
L^{\mu\nu}_{\rm FO} &=& A_\perp \tau_{\rm FO}^2 \int \!\!d\eta \Big({\cal E} + {\cal P}\Big) \times\cosh (\vartheta) \sinh (\vartheta)
\begin{bmatrix} 
0 & 0  & 0 & 1 \\ 
0 & 0  & 0 & 0 \\  
0 & 0  & 0 & 0 \\  
-1 & 0  & 0 & 0 \\
\end{bmatrix} =0\,.
\eeq
The second equality results from the symmetry of the integrand with respect to $\eta$ in the symmetric collision systems, which means ${\cal E}$ and ${\cal P}$ are $\eta$-even and $\vartheta$ is $\eta$-odd. Hence, the contribution to the total angular momentum coming from the orbital part is zero.
%
\subsection{Spin contribution}
\label{app:spincontri}
The spin angular momentum is given by
\beq
S^{\mu\nu}_{\rm FO} &=& \int
\Delta \Sigma _{\lambda } S^{\lambda,\mu\nu} = A_\perp \tau_{\rm FO} \int  d\eta \, U_{\lambda }^{\rm B} ~S^{\lambda,\mu\nu} ,
\label{SAM}
\eeq
where $S^{\lambda,\mu\nu}$ is given by Eq.~\rfn{eq:SGLW}. Using \rf{flow} one can easily calculate the components of $S^{\mu\nu}_{\rm FO}$. Assuming that the system's flow $\vartheta$ is $\eta$-odd the components  $S^{03}_{\rm FO}$ and $S^{12}_{\rm FO}$ 
\beq
S^{03}_{\rm FO}= -A_\perp \tau_{\rm FO} \int \!\! d\eta\, {\cal A}_3 C_{\kappa Z} \cosh (\vartheta)\,, \qquad
S^{12 }_{\rm FO}= -A_\perp \tau_{\rm FO} \int \!\! d\eta\, {\cal A}_1 C_{\omega Z} \cosh (\vartheta)\,,
\eeq
vanish only if $C_{\kappa Z}$ and $C_{\omega Z}$ are arbitrary $\eta$-odd functions or zero (one may check that the equations of motion preserve $\eta$ symmetry of these functions). The components
\beq
S^{13}_{\rm FO}&=& A_\perp \tau_{\rm FO} \int \!\! d\eta\, {\cal A}_3 \Bigg[ \LR \frac{{\cal A}_1}{{\cal A}_3} \cosh(\vartheta)\cosh(\Phi)-\frac{1}{2}\sinh(\vartheta)\sinh(\Phi) \RR C_{\omega Y}  + \frac{1}{4} \LR \sinh(\eta) +3 \sinh(\Phi+\vartheta)\RR C_{\kappa X} \Bigg],
\nn\\
-S^{01}_{\rm FO}&=& A_\perp \tau_{\rm FO} \int \!\! d\eta\, {\cal A}_3 \Bigg[ \LR \frac{{\cal A}_1}{{\cal A}_3} \cosh(\vartheta)\sinh(\Phi)-\frac{1}{2}\sinh(\vartheta)\cosh(\Phi) \RR C_{\omega Y}  + \frac{1}{4} \LR \cosh(\eta) +3 \cosh(\Phi+\vartheta)\RR C_{\kappa X} \Bigg].\nn
\label{spinFO}
\eeq
%
%
suggest that it is sufficient that $C_{\kappa X}$ and $C_{\omega Y}$ is $\eta$-odd and $\eta$-even, respectively, in order to have $S^{01}_{\rm FO}=0$ and $S^{13}_{\rm FO}\neq0$. In particular, at the initial time of evolution when $\vartheta=0$ one may choose Eq.~\rfn{spinini} for the initialization of the spin coefficients for the current work.

Similar reasoning can be given for $C_{\kappa Y}$ and $C_{\omega X}$ components, however, as in the case of $C_{\kappa Z}$ and $C_{\omega Z}$ they decouple from $C_{\kappa X}$ and $C_{\omega Y}$ dynamics, hence one can simply put them equal to zero.

One can also notice that the results obtained herein reduce to the results obtained for the special case of Bjorken flow~\cite{Florkowski:2019qdp}.
%
\section{General form of \texorpdfstring{$(\tilde{\omega}_{\mu \beta} p^\beta)^{\mbox{*}}$}{}}
\label{app:pol}
In this appendix we present the Lorentz transformed quantity $(\tilde{\omega}_{\mu \beta} p^\beta)^{\mbox{*}}$ for the non-boost invariant and transversely homogeneous system where all the spin coefficients ($C_{\boldsymbol{\kappa}}$ and $C_{\boldsymbol{\omega}}$) are present
\beq
(\tilde{\omega }_{0 \beta }p^{\beta })^{\mbox{*}} &=& 0\, ,\nn\\
(\tilde{\omega }_{1 \beta }p^{\beta })^{\mbox{*}} &=& m \alpha _p p_x \left[C_{\omega Z} m_T \sinh \left(y_p\right)-C_{\kappa Y} p_x \sinh (\Phi )+C_{\omega X} p_x \cosh (\Phi )+p_y \left(C_{\kappa X} \sinh (\Phi )+C_{\omega Y} \cosh (\Phi )\right)\right]\nn\\
&&+m_T \left[C_{\kappa Y} \sinh \left(\Phi -y_p\right)-C_{\omega X} \cosh \left(\Phi -y_p\right)\right]+C_{\kappa Z} p_y\, ,\nn\\
(\tilde{\omega }_{2 \beta }p^{\beta })^{\mbox{*}} &=& -m \alpha _p \Big[-p_y \left(C_{\omega Z} m_T \sinh \left(y_p\right)+p_x \left(C_{\omega X} \cosh (\Phi )-C_{\kappa Y} \sinh (\Phi )\right)\right)\nn\\
&& +m_T \left(m_T \cosh \left(y_p\right)+m\right) \left(C_{\kappa X} \sinh \left(\Phi -y_p\right)+C_{\omega Y} \cosh \left(\Phi -y_p\right)\right)+p_y^2 \left(-\left(C_{\kappa X} \sinh (\Phi )+C_{\omega Y} \cosh (\Phi )\right)\right)\Big]-C_{\kappa Z} p_x \,,\nn\\
(\tilde{\omega }_{3 \beta }p^{\beta })^{\mbox{*}} &=& -m \alpha _p \Big[p_x \left(C_{\omega X} \left(m_T \sinh \left(\Phi -y_p\right)+m \sinh (\Phi )\right)-C_{\kappa Y} \left(m_T \cosh \left(\Phi -y_p\right)+m \cosh (\Phi )\right)\right)\nn\\
&& +p_y \left(C_{\kappa X} \left(m_T \cosh \left(\Phi -y_p\right)+m \cosh (\Phi )\right)+C_{\omega Y} \left(m_T \sinh \left(\Phi -y_p\right)+m \sinh (\Phi )\right)\right)+C_{\omega Z} m_T \left(m \cosh \left(y_p\right)+m_T\right)\Big]\, .\nn\\
\eeq
This expression reduces to Eqs.~\rfn{OP3} for the physical scenario studied in this paper.
%
\section{Momentum averaged global spin polarization for the system with non-zero baryon chemical potential}
\label{app:withmu}
In this section we show the dependence of $\langle \pi_y \rangle$ component of the momentum averaged polarization on rapidity, transverse momentum, and azimuthal angle.
\begin{figure*}[t]
\centering
\includegraphics[width=8.6cm]{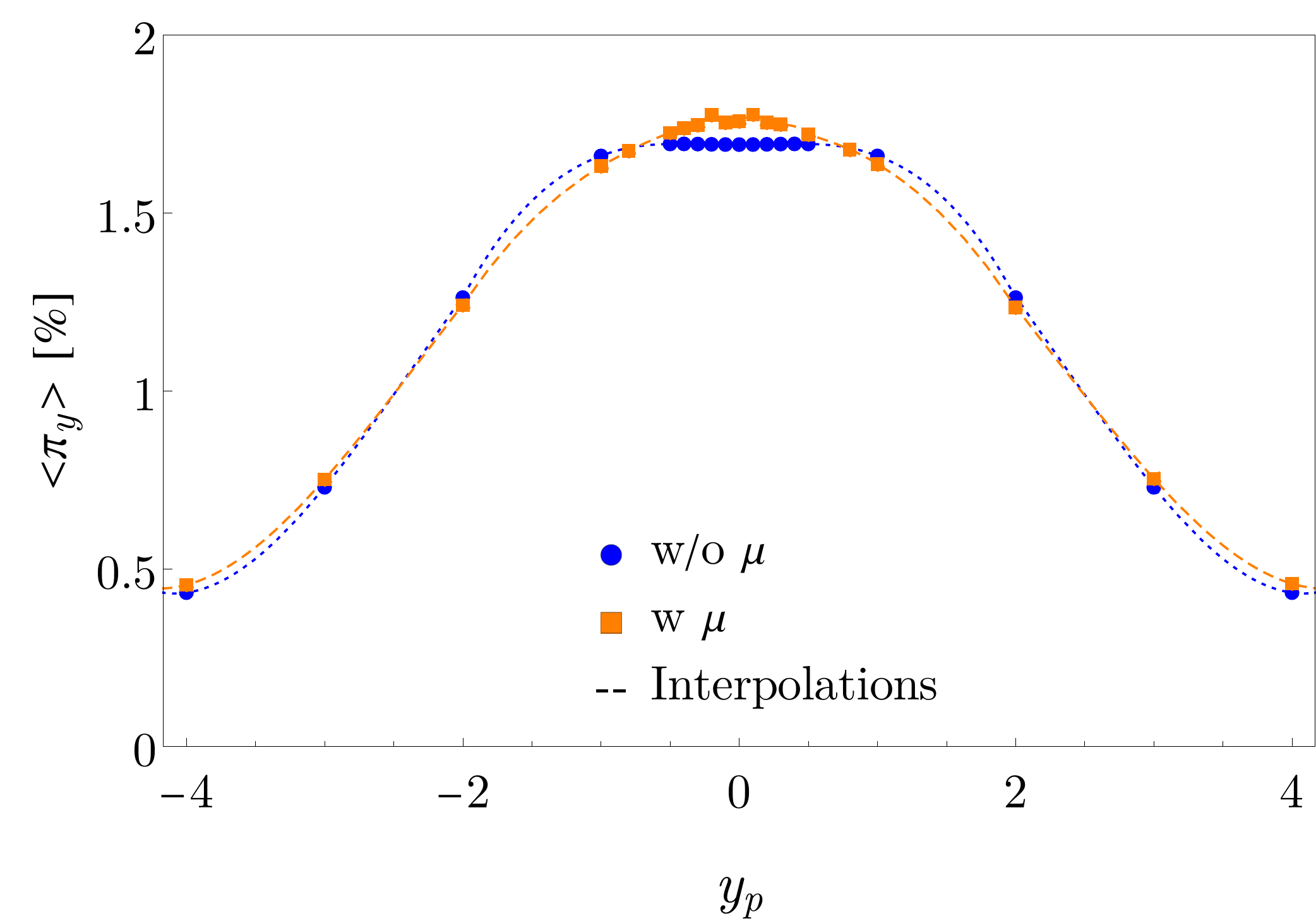}
\includegraphics[width=8.6cm]{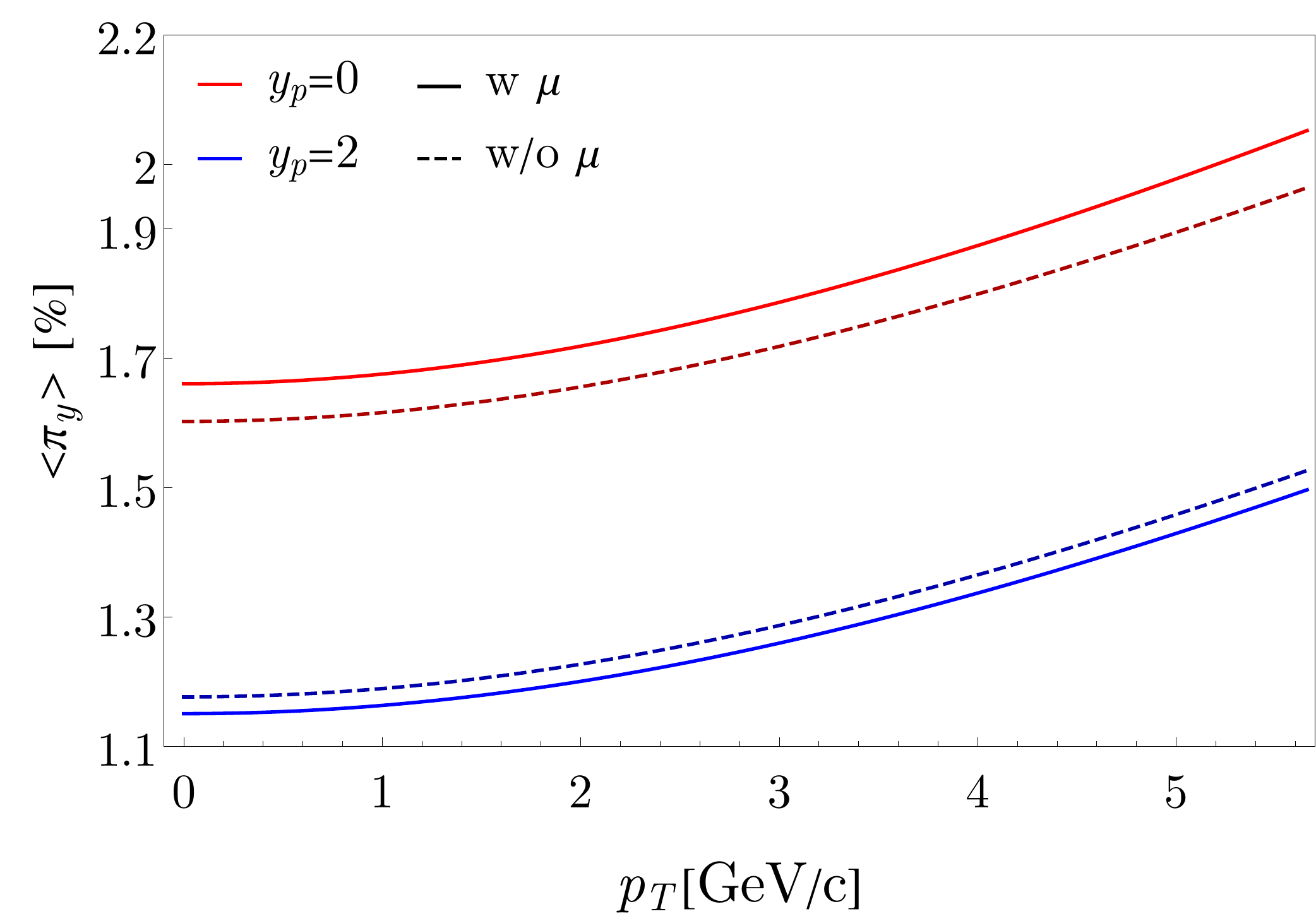}
\includegraphics[width=8.6cm]{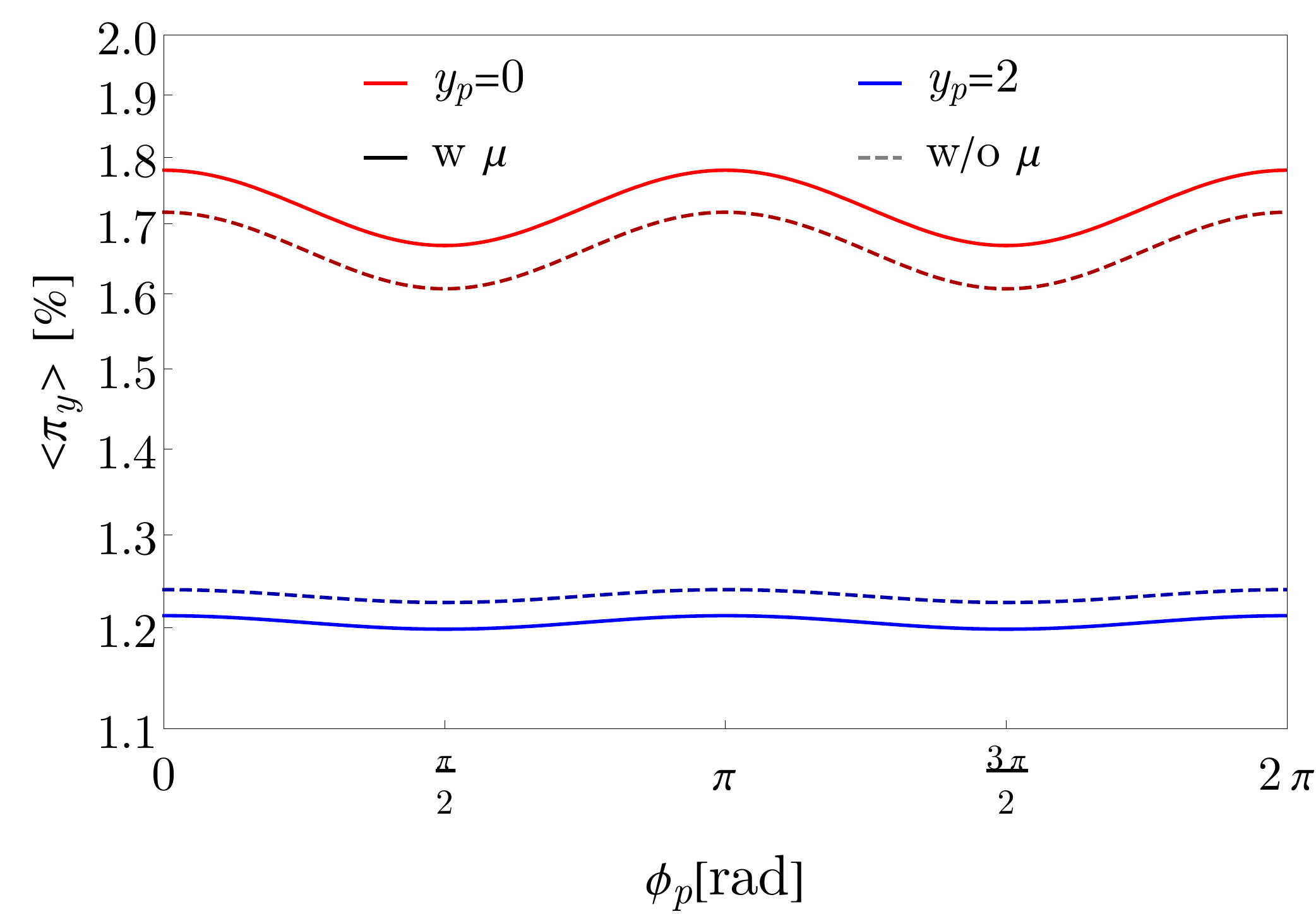}
\caption{$y_p$ (upper left),
$p_T$ (upper right), and $\phi_p$
(down) dependent $\langle \pi_y \rangle$ component of the momentum averaged polarization.}
\label{withMuplots1}
\end{figure*}
%
We note here that finite baryon chemical potential plays a role in the momentum averaged polarization behavior. One may notice that $\langle \pi_y \rangle$ behavior is more pronounced in midrapidity in comparison to baryon free system.
\end{widetext}
\bibliography{pv_ref}{}
\bibliographystyle{utphys}
\end{document}